\documentclass[fleqn,usenatbib,useAMS]{mnras}

\usepackage{graphicx}
\usepackage{amsmath}
\usepackage{amssymb}
\usepackage{xspace}

\usepackage{times}
\usepackage{multicol}

\newcommand{\Mch}{\ensuremath{{\rm M}_{\rm ch}}\xspace}

\newcommand{\Msun}{\ensuremath{{\rm M}_{\sun}}\xspace}

\newcommand{\iso}[2]{\hbox{${}^{#1}{\rm #2}$}\xspace}
\newcommand{\ud}{\rm {d}\xspace}
\newcommand{\Emax}{E_{\textrm{max}}\xspace}

\newcommand{\artis}{\textsc{artis}\xspace}
\newcommand{\cmfgen}{\textsc{cmfgen}\xspace}
\newcommand{\sumo}{\textsc{sumo}\xspace}

\newcommand{\keV}{\text{keV}\xspace}

\usepackage[T1]{fontenc}
\usepackage{ae,aecompl}

\usepackage{txfonts}

\title[Monte Carlo radiative transfer for nebular SNe Ia]{Monte Carlo radiative transfer for the nebular phase of Type Ia supernovae}

\author[Shingles et al.]{L. J. Shingles$^{1}$\thanks{Email: l.shingles@qub.ac.uk},
S. A. Sim$^1$,
M. Kromer$^{2,3}$,
K. Maguire$^1$,
M. Bulla$^{5,6}$,
C. Collins$^1$,\newauthor
C. P. Ballance$^4$,
A. S. Michel$^2$,
C. A. Ramsbottom$^4$,
F. K. R\"opke$^{2,3}$,
I. R. Seitenzahl$^{7}$,\newauthor
and N. B. Tyndall$^{4}$
\\
$^1$Astrophysics Research Center, School of Mathematics and Physics, Queen's University Belfast, Belfast BT7 1NN, Northern Ireland, UK\\
$^2$Zentrum f{\"u}r Astronomie der Universit{\"a}t Heidelberg, Institut f{\"u}r Theoretische Astrophysik, Philosophenweg 12, D-69120 Heidelberg, Germany\\
$^3$Heidelberger Institut f\"ur Theoretische Studien, Schloss-Wolfsbrunnenweg 35, D-69118, Heidelberg, Germany\\
$^4$Centre for Theoretical Atomic, Molecular and Optical Physics, School of Mathematics and Physics, Queen's University Belfast, Belfast BT7 1NN, Northern Ireland, UK\\
$^5$Oskar Klein Centre, Department of Physics, Stockholm University, SE-106 91, Stockholm, Sweden\\
$^6$Nordita, KTH Royal Institute of Technology and Stockholm University, Roslagstullsbacken 23, SE-106 91 Stockholm, Sweden\\
$^7$School of Science, University of New South Wales, Australian Defence Force Academy, Canberra, ACT 2600, Australia\\
}

\date{Accepted 2019 December 02. Received 2019 November 15; in original form 2019 October 02}
\pubyear{2019}

\begin{document}
\label{firstpage}
\pagerange{\pageref{firstpage}--\pageref{lastpage}}
\maketitle
\begin{abstract}
We extend the range of validity of the \artis 3D radiative transfer code up to hundreds of days after explosion, when Type Ia supernovae are in their nebular phase.
To achieve this, we add a non-local thermodynamic equilibrium (non-LTE) population and ionisation solver, a new multi-frequency radiation field model, and a new atomic dataset with forbidden transitions.
We treat collisions with non-thermal leptons resulting from nuclear decays to account for their contribution to excitation, ionisation, and heating.
We validate our method with a variety of tests including comparing our synthetic nebular spectra for the well-known one-dimensional W7 model with the results of other studies.
As an illustrative application of the code, we present synthetic nebular spectra for the detonation of a sub-Chandrasekhar white dwarf in which the possible effects of gravitational settling of \iso{22}{Ne} prior to explosion have been explored.
Specifically, we compare synthetic nebular spectra for a 1.06~\Msun white dwarf model obtained when 5.5 Gyr of very-efficient settling is assumed to a similar model without settling.
We find that this degree of \iso{22}{Ne} settling has only a modest effect on the resulting nebular spectra due to increased \iso{58}{Ni} abundance.
Due to the high ionisation in sub-Chandrasekhar models, the nebular [\ion{Ni}{ii}] emission remains negligible, while the [\ion{Ni}{iii}] line strengths are increased and the overall ionisation balance is slightly lowered in the model with \iso{22}{Ne} settling.
In common with previous studies of sub-Chandrasekhar models at nebular epochs, these models overproduce [\ion{Fe}{iii}] emission relative to [\ion{Fe}{ii}] in comparison to observations of normal Type Ia supernovae.
\end{abstract}


\section{Introduction}
\label{sec:intro}

Type Ia supernovae (hereafter SNe~Ia) have proven extremely valuable for cosmology as standardisable candles that enable measurement of the expansion history of the Universe \citep{Perlmutter:1999fb,Riess:1998hp}.
SNe~Ia are also a major source of Fe-group elements that are a crucial ingredient for galactic chemical evolution \citep{Nomoto:1984jh,Matteucci:1986ux}.
The broad consensus is that SNe~Ia involve the thermonuclear destruction of an electron-degenerate white dwarf \citep[WD;][]{Hoyle:1960bk,Bloom:2012he}, with their extremely high peak brightness and subsequent decline powered by the radioactive decay of \iso{56}{Ni} produced in the explosion \citep{Arnett:1979hg}.
However, despite decades of observational and theoretical studies \citep[for a review, see e.g.][]{Hillebrandt:2013hy}, even the most typical SN Ia events are still poorly understood at a fundamental level, with no conclusive determination of either the progenitor systems nor the mechanism by which they explode.
A recent review of explosion models is given by \citet{Livio:2018ik}.

A natural candidate for triggering the ignition is for the WD to approach the Chandrasekhar mass (\Mch) limit by accretion from a companion star.
A near-\Mch C-O WD that undergoes a pure detonation would result in a composition of mostly Fe-group elements \citep{Arnett:1969gf,Hansen:1969eu}, which would fail to explain the observed lines of intermediate-mass elements such as silicon and sulphur in SN Ia spectra.
Alternatively, a deflagration, which proceeds sub-sonically, would allow the stellar material to expand ahead of the flame front and reduce the densities at which the burning takes place.
Although pure-deflagration models produce intermediate-mass elements, self-consistent three-dimensional models \citep[e.g.,][]{Fink:2014gg} fail to match the spectral features and high luminosities observed in normal SNe~Ia.
The limitations of these two extremes have motivated \Mch models with a delayed detonation: a detonation that is preceded by a pre-expansion phase due to subsonic deflagration burning \citep{Khokhlov:1991tl}.
For recent multi-D realisations of this model see e.g., \citet{Kasen:2009ff} and \citet{Seitenzahl:2013fz}.

Another way for nuclear burning to take place at lower densities is for the detonation to occur in a WD with a mass below the Chandrasekhar mass limit (a sub-\Mch explosion).
Finding a trigger for the explosion has proved challenging, although a promising explanation is that accretion from a companion triggers a surface detonation in a thin He-shell, which then leads to a second detonation of the C-O core (the double-detonation scenario; \citealt{Livne:1990iz,Livne:1990cw}).

One difficulty in determining the progenitor systems and explosion mechanisms of SNe~Ia is that a variety of distinct scenarios lead to very similar predictions for the observable light curves and spectra.
For example, \citet{Ropke:2012im} compared observations of the well-studied SN2011fe with two different explosion models, the N100 DDT (deflagration-to-detonation transition) model by \citet{Seitenzahl:2013fz} and a model of a violent merger between 1.1 \Msun and 0.9 \Msun WDs \citep{Pakmor:2012ds}.
For at least the first 30 days, the light curves and spectra calculated for the two explosion scenarios both produced a reasonable fit to the observations, with neither being clearly excluded or favoured by the data.

A promising resolution to the degeneracy between scenarios is to observe supernovae at very late times, when they have reached their nebular phase and become optically thin at most wavelengths.
In SNe~Ia, the transition to a nebular phase occurs around several tens of days after explosion, after which photons emitted by the gas typically escape rather than being reabsorbed by the ejecta.
Under these conditions, excitation by both radiative and collisional processes are slow, and only lowest-energy states have significant populations.
The spectra of nebular-phase SNe~Ia are therefore dominated by forbidden line emission from the lowest-lying metastable states of Fe, Co, and Ni \citep{Spyromilio:1992ca}.

A consequence of the direct relationship between velocity and radius under homologous expansion is that nebular emission lines and their Doppler shifts reveal the distribution of emitting gas throughout the entire ejecta \cite[for a review, see][]{Jerkstrand:2017gj}.
This includes the inner core where ignition has taken place and for which some simulations predict an asymmetric three-dimensional structure, dependent on the degree of mixing that has taken place following ignition and prior to the homologous expansion phase.
For example, detections of double-peaked line profiles of Fe and Co in nebular-phase spectra have been claimed \citep{Dong:2015hs}, suggesting that explosions result in bimodally-structured ejecta \citep[see also][]{Mazzali:2018gv}.

The first models of the SN~Ia nebular phase were the one-zone models of \citet{Axelrod:1980vk}, who applied a simplified treatment of gamma-ray absorption and a high-energy limit approximation to non-thermal ionisation and heating processes.
Since the work of \citet{Axelrod:1980vk}, SNe~Ia nebular studies have advanced to multi-zone models with more detailed treatment of gamma-ray transfer, radiative processes, and non-thermal physics \citep{RuizLapuente:1992iu,Liu:1997hs,Mazzali:2001gz,Hoflich:2004fka,Kozma:2005hz,Li:2012dx,Botyanszki:2017hx}.
Such studies have demonstrated the power of nebular phase observations to constrain key properties of the inner ejecta and test the accuracy with which different explosion scenarios can reproduce composition, ionisation and thermal profiles consistent with observations.

However, despite the effectiveness of late-time spectra in constraining the composition and geometry of the inner ejecta, full three-dimensional radiative transfer simulations for state-of-the-art SNe~Ia explosion models in the nebular phase are rare in the literature: many existing studies that depart from spherical symmetry are based on simplified toy geometries \citep[e.g.,][]{Maeda:2010fx} or superposition of 1D models \citep[e.g.,][]{Mazzali:2018gv} and/or assume optically-thin emission \citep[e.g.,][]{Botyanszki:2018kr}.

Our aim is therefore to develop and present numerical spectrum synthesis for sets of modern, multi-D explosion simulations in order that they can be used to constrain explosion theories.
Accordingly, in this paper, we present extensions to the 3D Monte Carlo radiative transfer code \artis in order to extend its validity to the modelling of the nebular phase.

\section{Numerical Method}
\label{sec:method}

We model radiative transfer in SNe~Ia using the \artis code described by \citet{Sim:2007is} and \citet{Kromer:2009hv} \citep[see also][for a description of the method]{Lucy:2005cx}.
\artis is a three-dimensional Monte Carlo radiative transfer code that uses indivisible energy packets \citep{Lucy:2002hw}.
The code has been extended to trace polarisation by \citet{Bulla:2015dt}.
Here, we describe further developments to the code to extend its validity into the nebular epoch.

The \artis code solves for the plasma conditions at each time-step by neglecting time-dependent terms and assuming that the plasma is in statistical, thermal, and ionisation equilibrium.
These conditions imply equilibrium level populations (as rates into and out of each level are balanced), equilibrium electron temperature (as heating is balanced by cooling, provided a solution falls within the allowed temperature range), and equilibrium ionisation balance (as recombination is balanced by ionisation), respectively.

The steady-state assumption has been shown to remain valid for supernovae well into the nebular phase to very late times after the explosion.
In the Type II supernova SN1987A, \citet{Fransson:1993gq} detected a freeze-out of the ionisation state due to slowing of ionisation and recombination at around $\sim$ 800 days.
More relevant for our study of SNe~Ia are the radiative transfer models of SN2011fe by \citet{Fransson:2015ct}.
Their comparison between a steady-state model and one in which time-dependent effects have been included show that the time-dependency starts to become important at around $\sim$ 700 days after explosion.
With these results as a guide, we limit our use of the \artis code to times earlier than $\sim$ 700 days at which the steady-state assumption is a reasonable approximation.

\subsection{New atomic dataset}
\label{sec:atomicdata}

Prior calculations with \artis have used the `big\_gf-4' atomic dataset described by \citet{Kromer:2009hv}, with transition line data from \citet{Kurucz:1995ui} and \citet{Kurucz:2006iz}.
This earlier atomic dataset did not include sufficiently complete data (such as electron collision and radiative transition rates) to accurately treat forbidden transitions.
However, the nebular spectra of SNe~Ia are dominated by forbidden lines, and therefore we have adopted a new atomic dataset for use in the studies presented here.

Our dataset is based on the atomic data compilation of \cmfgen\footnote{Available at \url{http://kookaburra.phyast.pitt.edu/hillier/web/CMFGEN.htm}} \citep{Hillier:1990ux,Hillier:1998gw}, with some modifications described below.
Our standard models in this work include \ion{C}{i-iv}, \ion{O}{i-iv}, \ion{Ne}{i-iii}, \ion{Mg}{i-ii}, \ion{Si}{i-iv}, \ion{S}{i-iv}, \ion{Ar}{i-iv}, \ion{Ca}{i-iv}, \ion{Fe}{i-v}, \ion{Co}{ii-iv}, and \ion{Ni}{ii-v}.

Energy levels in the \cmfgen compilation are generally sourced from \citet{Kurucz:1995ui}, \citet{Kurucz:2006iz}, and the NIST Atomic Spectra Database\footnote{Available at \url{http://physics.nist.gov/asd}}.
The photoionisation cross sections are sourced from the Opacity Project \citep{Seaton:1987do} and the Fe project \citep{Hummer:1993tf}, which is also a source of electron collisional data.
Forbidden line data is sourced from \citet{Quinet:1996uj} for \ion{Ni}{ii}, \citet{Garstang:1958fz} for \ion{Ni}{iii}, \citet{Quinet:1996wf} for \ion{Fe}{ii}, and \citet{Quinet:1996tw} for \ion{Fe}{iii}.

Our photoionisation cross sections are scaled such that the total recombination rate of each ion when LTE level populations at 6000 K are assumed exactly matches the tabulated recombination rates of Nahar \citep{Nahar:1997tj,Nahar:1994ft,Nahar:1996jp,Nahar:1997bn,Nahar:1998bo,Nahar:2001im} for Fe and Ni$^{2{+}}$, and CHIANTI \citep{Dere:1997gd,DelZanna:2015ie} for all other elements/ions.
The recombination rate coefficients of the individual levels (derived from the photoionisation cross sections by the Milne relation) are then applied to the specific non-LTE populations during the simulation.
The use of non-LTE populations has a very small effect on the ion-recombination rates, since recombination typically proceeds from states in the ground multiplet, which remain close to their LTE populations.
The CHIANTI recombination rates for Co ions are obtained by adding fits to the radiative rate from \citet{Landini:1991tz} to a fitted dielectronic recombination rate from \citet{Mazzotta:1998hi}.

In addition to the \cmfgen database, we use updated collision strengths for Co$^{{+}}$ from \citet{Storey:2016bv}, and Co$^{{+}}$ photoionisation cross sections, Co$^{2{+}}$ energy levels and bound-bound transition rates from \citet{Tyndall:2016gv}.

To calculate bound-bound collisional transition rates, we apply effective collision strengths for the typical electron temperature ($\sim 6 \times 10^3$ K) where this data is available (e.g., Co$^{{+}}$ and most ions in the \cmfgen compilation including Fe$^{0{+}}$--Fe$^{3{+}}$).
For collision strengths that are specified only between LS terms without J-splitting \citep[such as Ni$^{2{+}}$ data by][]{Watts:1999io}, we apply the collisional strength to each relevant pair of J-specific levels with a factor accounting for the fraction of statistical weight in the J-specific upper level relative to the total statistical weight of the upper term.
Collisional rates for radiatively-permitted transitions without effective collision strengths are estimated with the \citet{vanRegemorter:1962gf} approximation.
For forbidden transitions without collision strengths in the database, we adopt an effective collision strength of $\Upsilon_{jk} = 0.01 g_j g_k$ where $g_j$ and $g_k$ are the statistical weights of the lower and upper levels (similar to \citealt{Axelrod:1980vk}).
Despite these improvements, it should be noted that the accuracy of high-quality simulations is limited by the quality of the atomic data.
In particular, more-complete sets of collision strengths (e.g., for Fe$^{3{+}}$ and higher, and Co) and improvements to the photoionisation cross sections are priorities for future work.

\subsection{Radiation field and photoionisation estimators}
\label{sec:radiationfield}
To calculate the rates of radiative excitation, photoionisation, and bound-free heating, the most accurate technique is to directly count the contribution along every Monte Carlo packet flight path.
However, this requires storing a rate estimator for each atomic process in every grid cell (or alternatively, storing the full history of all packets), which does not easily scale to the large grid sizes of 3D models.
Consequently, we prioritise accurate treatment of photoionisation, and currently adopt a radiation field model for other atomic processes (see below).

\subsubsection{Photoionisation estimators}
We use the full packet trajectories to obtain estimators for all photoionisation transitions \citep[using equation 44 of][]{Lucy:2003bz}.
%, and to construct a model of the radiation field intensity as a function of frequency.
This provides detailed level by level photoionisation rate coefficients, which are then used in our NLTE solution (Section \ref{sec:nltepopulationsolver}).

\subsubsection{Bound-bound and heating estimators}
To estimate rate coefficients for bound-bound process (as required for our NLTE solution -- see Section \ref{sec:nltepopulationsolver}) and radiative heating rates, we construct a radiation field model in each grid cell.

In prior versions of the \artis code, the radiation field in each grid cell is modelled as a dilute blackbody described by two parameters: the radiation temperature $T_R$ and the dilution factor $W$, by which the radiation field has been weakened relative to the Planck function.
The radiation temperature is chosen such that the mean frequency of the Planck function is equal to the energy-weighted mean frequency of the propagating packets (\citealp{Mazzali:1993vl}, also used by \citealp{Long:2002fs}).
The dilution factor is then used to scale the blackbody field to the required energy density.
The mean intensity and frequency moments are obtained from estimators calculated with the packet trajectory summation technique described by \citet{Lucy:1999wd}.

In the nebular epoch, however, the radiation field within the SN Ia ejecta will deviate substantially from a blackbody, and so we require a more detailed treatment of the radiation field.
%Although it is possible to account for the estimated flux in each spectral line by summing over packet trajectories (in the 1D models present here), the large number of estimators required for this method would consume too much memory when \artis is applied to three-dimensional models (with $\sim50^3$ cells).
%So, to allow for a more-precise estimation of the radiation field
To achieve this with reasonable memory requirements, we accumulate the mean intensity $J$ and the moment $\nu J_\nu$ estimators independently within each of a set of frequency bands.
Specifically, for each segment of a packet trajectory, we calculate a contribution to the $J_i$ and $\nu J_i$ estimators of frequency band $i$ that contains the packet's co-moving frequency $\nu$, i.e.,
\begin{align}
	\Delta J_i &= \frac{\Delta r \, nh\nu}{\Delta V \Delta t}\\
	\Delta (\nu J_i) &= \nu \Delta J_i,
\end{align}
where $\Delta r$ is the distance travelled by a packet representing $n$ photons with co-moving frequency $\nu$ during the time $\Delta t$ in a grid cell with a volume of $\Delta V$.
The ratio of $\nu J_\nu$ to $J_\nu$ gives the intensity-weighted mean photon frequency ($\bar{\nu}$) for the frequency band.
We then assign to the band a radiation temperature, $T_{R,i}$, obtained by iteratively searching for a temperature in the range 500 to 250,000 K such that the windowed Planck function has the closest matching $\bar{\nu}$ to the estimator value.
A dilution factor $W_i$ is then calculated to scale the intensity of the windowed Planck function to the estimator value.
A similar binning approach was adopted by the Monte Carlo code of \citet{Higginbottom:2013ja}.

In this work, we fit dilute blackbody functions to 255 bins from 1085 to 40,000 \AA. 1085 \AA ~has been chosen to ensure that recombination to the highest metastable level of Fe$^{{+}}$ has a threshold frequency outside the bins. Short-ward of 1085 \AA ~(where packet statistics are limited and opacities are relatively high), we adopt the local electron temperature for the radiation field but impose a scaling factor to enforce energy conservation.

With the radiation field model obtained in this way, we can then estimate bound-bound radiative rates using equation 10 of \citet{Lucy:2003bz} with the blue-wing mean intensity estimated from the appropriate bin of the radiation field model. This approach is clearly less accurate that using detailed line-by-line estimators (i.e. equation 46 of \citealt{Lucy:2003bz}) however it is substantially less demanding of memory requirements and, as discussed by \cite[][see their fig 14]{Kerzendorf:2014gv}, adopting mean intensities from a radiation field model will often provide a good estimate. We also use the radiation field model to determine bound-free heating rates for use in our calculation but we note that heating via non-thermal deposition is always dominant in the calculations we discuss here.
%We then use this radiation field model to estimate the rates of radiative excitation and bound-free heating.
%
%Using a detailed radiation field is particularly important for the accuracy of radiative transition rates.

\subsection{Non-LTE ionisation and population solver}
\label{sec:nltepopulationsolver}
We calculate the ionisation balance and level populations by solving the equations of statistical equilibrium for each grid cell and time-step.
For each element, we solve for the populations of all non-local thermodynamic equilibrium (NLTE) energy levels of all included ionisation stages simultaneously.
The simultaneous whole-element solution enables the easy implementation of multiple ionisations from the non-thermal solver (Section \ref{sec:ntelectrons}).

In matrix form, the statistical balance equations are written
\begin{equation}
	\sum_{j \ne k}{n_{j} R_{j\rightarrow k} - n_{k} \left(\sum_{j \ne k} R_{k\rightarrow j}\right)} = 0
\end{equation}
for each pair of levels $j$ and $k$, where $n_j$ and $n_k$ are the populations of level $j$ and $k$, and $R_{j\rightarrow k}$ is the total rate per population of all processes that remove population from level $j$ and add it to level $k$.

Treating all levels of all ions in NLTE has a large memory and computational cost, so we restrict the number of levels treated in full NLTE.
With the first $l$ levels treated in full NLTE, we combine the population of the levels $l+1\ldots l_\mathrm{max}$ into a `super-level' \citep{Anderson:1989dl}.
While the super-level is treated as an additional NLTE level that can vary in population, the ratios between the populations of the levels that comprise it are calculated with the assumption that these levels are in Boltzmann equilibrium with each other at the electron temperature.
For most ions, we treat the first 80 levels in NLTE, but increase this to 197 NLTE levels for Fe$^{{+}}$ to ensure coverage of all metastable levels (those with no permitted transitions to the ground state).

The processes contributing to the rate equation include excitation and de-excitation by radiation and collisions with thermal and non-thermal electrons.
We also include radiative and collisional recombination, and ionisation via photoionisation and collisions with both thermal and non-thermal electrons.
The resulting matrix equation is numerically solved with the gsl\_linalg\_LU\_solve function of the GNU Scientific Library\footnote{Available at \url{http://www.gnu.org/software/gsl/}} \citep{Gough:2009tb}.

\subsection{Treatment of non-thermal energy deposition}
\label{sec:ntelectrons}

The majority of the emission from SNe~Ia is powered by energy injection from nuclear decays \citep{Colgate:1969jw}.
The decays of \iso{56}{Ni}, \iso{56}{Co}, and \iso{48}{V} produce $\gamma$-rays, which deposit energy into the plasma via Compton scattering with free and bound electrons.
A fraction of these decays also produce positrons, which deposit their energy locally \citep{Axelrod:1980vk} and we assume that the positron deposition takes place within the same grid cell as the decaying nuclei.
We also assume that all deposited energy is injected uniformly at a constant rate within each grid cell during each time-step.

We calculate the total (high-energy) deposition rate density in a grid cell by adding the estimated absorption rate of $\gamma$-ray packets to the positron contributions from the decays of \iso{56}{Co} and \iso{48}{V}. Using the Bateman equation for the radioactive decay chain $\iso{56}{Ni} \xrightarrow{\tau = 8 d} \iso{56}{Co} \xrightarrow{\tau = 113 d} \iso{56}{Fe}$, the number density of \iso{56}{Co} nuclei is
\begin{equation}\label{eqn:nco56}
	n_\mathrm{Co56}(t) = \frac{\lambda_\mathrm{Ni56}}{\lambda_\mathrm{Co56} - \lambda_\mathrm{Ni56}} \left(e^{-\lambda_\mathrm{Co56}t} - e^{-\lambda_\mathrm{Ni56}t}\right) \frac{\rho(t) X_{0,\mathrm{Ni56}}}{M_\mathrm{Ni56}},
\end{equation}
where $\lambda_\mathrm{Ni56}$ and $\lambda_\mathrm{Co56}$ are inverse mean-lifetimes ($\tau^{-1}$) of \iso{56}{Ni} and \iso{56}{Co}, $\rho(t)$ is the mass density at time $t$, $X_{0,\mathrm{Ni56}}$ is the initial mass fraction of \iso{56}{Ni} in the cell, and $M_\mathrm{Ni56}$ is the mass of a \iso{56}{Ni} nucleus. The number of \iso{56}{Co} decays per unit time, per unit volume is obtained by multiplying Equation (\ref{eqn:nco56}) by $\lambda_\mathrm{Co56}$.
While 81\% of \iso{56}{Co} decays proceed via electron capture, the remaining 19\% proceed via emission of a positron with an energy in the range 0--1.459 MeV, and a mean energy of 610 \keV.
Hence, the energy deposited per unit time, per unit volume due to positrons emitted in \iso{56}{Co} decays is given by
\begin{equation}
	D_{e^+, \mathrm{Co56}} = 0.19 \cdot 610\ \keV \cdot \lambda_\mathrm{Co56}\ n_\mathrm{Co56}.
\end{equation}
Similarly for the decay of \iso{48}{V}, which proceeds 49\% of the time with the emission of positrons having a mean energy of 290 keV,
\begin{equation}
	D_{e^+, \mathrm{V48}} = 0.49 \cdot 290\ \keV \cdot \lambda_\mathrm{V48}\ n_\mathrm{V48}.
\end{equation}
The total energy deposition rate then is the sum of positron and gamma-ray deposition components,
\begin{equation}\label{eq:deprate}
	D_{\mathrm{tot}} = D_\gamma +  D_{e^+, \mathrm{Co56}}  + D_{e^+, \mathrm{V48}},
\end{equation}
where $D_\gamma$ is rate of deposition due to gamma ray absorption.

At the low densities of SN Ia ejecta several hundred days after explosion, the high-energy electrons and positrons resulting from the energy deposition are not effectively thermalised, leading to a population of high-energy leptons with a non-thermal distribution of energies.
As the high-energy particles slow down and secondary electrons are produced by ionisations, the resulting electron energy spectrum contributes a high-energy tail to the Maxwellian distribution of the thermal electrons.
These non-thermal particles contribute to the ionisation and excitation of ions, and heating of the thermal electrons.

To treat the energy deposition as merely a heating source for the thermal pool of electrons (which is a good approximation at early times) is inadequate for the nebular phase, during which the ionisation balance is largely controlled by collisions between ions and non-thermal electrons \citep{Kozma:1992cy}.
We therefore require a treatment of the ionisation and heating caused by non-thermal electrons.

We use a detailed treatment to account for the contributions to heating, excitation, and ionisation from non-thermal electrons by first calculating the non-thermal energy distribution using the Spencer-Fano equation \citep{Spencer:1954cb} as explained below.

\subsection{The non-thermal degradation equation}
\label{sec:ntdegradation}

The Spencer-Fano equation is a specific case of the Boltzmann equation that balances the number of electrons entering and leaving each energy interval.
Inside supernova ejecta, the sources and sinks are the energy injected by $\gamma$-rays and positrons, excitation and ionisation of ions, and Coulomb scattering with thermal electrons.

The calculated deposition rate from \autoref{eq:deprate} is used to scale a source function ($S(E)$, where $E$ is the energy) with a top-hat profile across the narrow range of energies ($\sim$15.5--16 \keV).
The particular energy at which we apply the non-thermal deposition does not alter the solution significantly, provided it is high enough for the ionisation rates to converge.
Our testing found that the ionisation rates of Fe ions converged to a tolerance of about 10\% by increasing the injection energy above 16 keV.

We numerically solve an integral form of the Spencer-Fano equation similar to \citet{Kozma:1992cy} but with an additional term to account for Auger electrons released from ionisations of inner-shells,
\begin{equation}\label{eqn:sf}
\begin{split}
&\sum_{j} n_{j} \sum_{k} \int_{E}^{E+E_{j\rightarrow k}} y(E') \sigma_{\mathrm{exc},j\rightarrow k} (E')dE' +  y(E) L_e(E)\\
&+ \sum_i N_i \sum_m \int_{E}^{\Emax} y(E') \int_{E'-E}^{(E'+E)/2} \sigma_{\mathrm{ion},m}(E',\epsilon) d\epsilon dE'\\
= &\sum_i N_i \sum_m \int_{2E+I_m}^{\Emax} y(E') \int_{E+I_m}^{(E'+I_m)/2}\sigma_{\mathrm{ion},m}(E',\epsilon) d\epsilon dE'\\
&+\sum_i N_i \sum_m \int_{E}^{\Emax} \delta(E' - \bar{E}_{\mathrm{Auger,}m}) dE' \int_{E}^{\Emax} y(E') \sigma_{\mathrm{ion},m}(E') dE'\\
&+ \int_{E}^{\Emax} S(E')dE',
\end{split}
\end{equation}
where summation runs over ions ($i$), energy levels ($j$ and $k$), and electron shells ($m$), $n_{j}$ is the population density of level $j$, $E_{j\rightarrow k}$ and $\sigma_{\mathrm{exc},j\rightarrow k}$ are the energy difference and cross section of the excitation transition from level $j$ to $k$, $N_i$ is the population density of ion $i$, $\sigma_{\mathrm{ion},m}$, and $I_m$ are the impact ionisation cross section and ionisation potential of electron shell $m$, $L_e$ is the loss function for Coulomb interactions with thermal electrons \citep[which we calculate identically to][]{Kozma:1992cy}, and $\Emax$ is the maximum energy up to which the solution is defined.
The solution to Equation (\ref{eqn:sf}) is the energy degradation function $y(E) = \ud f/\ud E$, where $f$ is the electron number flux.
Thus, $y$ is a distribution function for the flux of non-thermal particles.

Similar to \citet{Li:2012dx}, we use the electron-impact ionisation cross section ($Q_{\mathrm{ion},m}$) fitting formula of \citet{Younger:1981bq}, with data from \citet{Arnaud:1985wq} and \citet{Arnaud:1992ft}.
To obtain the differential cross sections, we estimate the energy distribution of ejected electrons with the formula of \citet{Opal:1971cp},
\begin{equation}
	P(E_p, E_s) = \frac{1}{J \arctan[(E_p - I_m) / 2 J_m]} \frac{1}{1 + (E_s/J_m)^2},
\end{equation}
where $I_m$ is the ionisation potential of shell $m$, $E_p$ and $E_s$ are the energies of the primary and secondary electrons, and $J_m$ is a fitting parameter that acts as a cut-off energy for secondary electrons.
Following \citet{Kozma:1992cy}, we use $J_m=24.2$ eV for \ion{Ne}{i}, and $J_m=10.0$ eV for \ion{Ar}{i}, as measured by \citet{Opal:1971cp} and $J_m = 0.6I_m$ for all other ions. The differential cross section is then
\begin{equation}
\sigma_{\mathrm{ion},m}(E, \epsilon) = Q_{\mathrm{ion},m}(E) P(E, \epsilon - I_i),
\end{equation}
where $\sigma_{\mathrm{ion},m}$ is the total cross section, and $\epsilon$ is the kinetic energy of the secondary electron.

We have implemented the capability to include excitation of bound electrons by non-thermal collisions for a subset of the permitted lines, making use of the \citet{vanRegemorter:1962gf} approximation with a Gaunt factor estimated from the first two terms of the fitting formula given in Equation 5 of \citet{Mewe:1972vw}.
However, this part of the simulation is particularly computationally demanding, and consequently this part of the implementation is not used in the initial simulations presented here.

The solution to Equation (\ref{eqn:sf}) at energy $E$ only depends on quantities evaluated at energies between $E$ and $\Emax$.
This means that when we discretise the integrals, the resulting set of linear equations forms an upper-triangular matrix that is easily solved on a computer.
The solution vector then contains the electron degradation spectrum $y$ evaluated on our grid of energy points.

\subsection{Non-thermal ionisation rates}
\label{sec:ntrates}

With a known electron degradation spectrum, the fraction of deposition energy going into ionisation of electron shell $s$ of ion $i$ is obtained from
\begin{equation}\label{eqn:etaionisation}
	\eta_s = \frac{N_i I_s}{E_{\rm{init}}} \int_{I_s}^{E_{\rm{max}}} \sigma_s (E) y(E) \,\ud E,
\end{equation}
where $I_s$ and $\sigma_s$ are the ionisation potential and impact ionisation cross sections of shell $s$.
Similarly, the non-thermal ionisation rate of ion $i$ is given by
\begin{equation}\label{eqn:ionisationrate}
	\Gamma_i N_i = \frac{N_i \epsilon_{\rm{dep}}}{E_{\rm{init}}} \sum_s{ \int_{I_s}^{E_{\rm{max}}}{\sigma_s (E) y(E) \,\ud E}}.
\end{equation}

When a non-thermal electron impact frees an electron from an inner shell, the relaxation of the resulting ion can eject further bound electrons (the Auger effect).
In our standard \artis models, we use the probabilities of ejecting one or two Auger electrons given by
\citet{Kaastra:1993uk} to calculate rates of double- and triple-ionisation in the non-LTE population/ionisation solver\footnote{I.e.,
in the solver, the ionisation rate is used to connect the target ion to the ground states of the species with one degree higher ionisation (rate proportional to probability of no Auger electron), two degrees higher (proportional to rate for one Auger electron, etc.)}.
Note, however, that we do not currently follow the photons produced in the refilling of inner shells, but we do include an extra term in our form of the Spencer-Fano equation to allow Auger electrons to contribute to heating, ionisation, and excitation.

\section{Verification of method}
\label{sec:verification}

In this section, we present the results of several calculations made to test and demonstrate the newly implemented code features.
We first show the results of an idealised test of the non-thermal solver (Section \ref{sec:nt-solver}) in which only a single element is included.
We then (Section \ref{sec:w7run}) present and discuss a full  spectrum synthesis calculation for the well-known W7 model \citep{Nomoto:1984jh,Iwamoto:1999jd}.

\subsection{Non-thermal solver}
\label{sec:nt-solver}
\begin{figure}
 \begin{center}\includegraphics[width=\columnwidth]{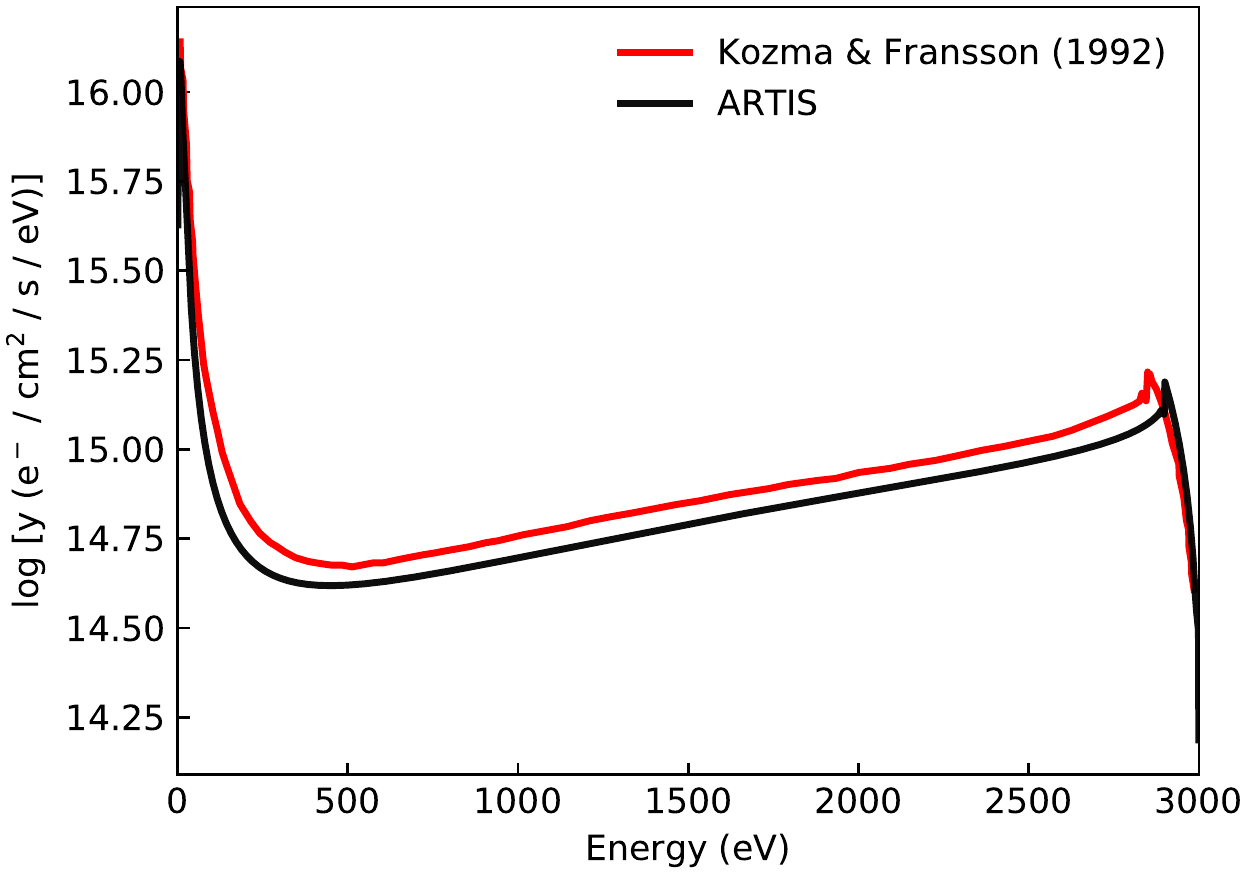}\end{center}
 \caption{Electron degradation function, $y$ for a pure-O plasma using \artis (black).
With the same thermal electron density ($10^8$ cm$^{-3}$) and ionisation fraction ($10^{-2}$), and a similar source function (spread across 2.9-3.0 keV), the \artis result is similar to \citet{Kozma:1992cy}.
The red line has been digitised from figure 1 of \citet{Kozma:1992cy}. \label{fig:ntspecpureoxygen}}
\end{figure}

To verify our implementation of the Spencer-Fano solver, we calculate the electron degradation function for a pure-O plasma with the same ionisation fraction ($0.01$) and free electron density ($10^8$ cm$^{-3}$) as \citet{Kozma:1992cy}.
The degradation function is shown in \autoref{fig:ntspecpureoxygen}, with a digitised version of the degradation function in figure 1 of \citet{Kozma:1992cy} for comparison.
We find close agreement, with a small difference that is likely due to differences in the chosen source function, i.e. the distribution of energies at which we inject the non-thermal energy.

\subsection{W7 calculation}
\label{sec:w7run}

To more fully test the nebular phase capabilities of the code, we have carried out test calculations for the nebular spectrum at 330 days (post explosion) for the W7 model of \citet{Nomoto:1984jh} with the nucleosynthesis of \citet{Iwamoto:1999jd}.
The W7 model is derived from a 1D simulation of the deflagration of a \Mch C-O WD with the deflagration speed having been chosen to yield an overall good level of agreement with the known properties of normal SNe~Ia.
We adopt this model for our tests since it is well known in the literature and has been widely used in previous studies \citep{RuizLapuente:1995cv,Liu:1997hs,Sollerman:2004fy}.
In particular, adopting this model (and epoch) allows us to directly compare with the existing nebular phase spectrum synthesis calculations made with the \sumo code (see Section \ref{sec:w7comparison}).
It should be noted, however, that this version of the W7 model is not fully representative of modern Chandrasekhar mass models.
For example, revisions to the nucleosynthesis yield calculations \citep{Leung:2017tz,Nomoto:2018iw} involve significant updates for some of the isotopes relevant to the nebular phase.
More importantly, the 1D symmetry of the model is not well-justified based on modern multi-dimensional explosion simulations \citep[e.g.,][]{Seitenzahl:2013fz,Sim:2013dy}.
We will address these questions in subsequent studies in which modern multi-dimensional explosion simulations will be used.

\subsubsection{Ionisation and thermal structure}

\begin{figure}
 \begin{center}\includegraphics[width=\columnwidth]{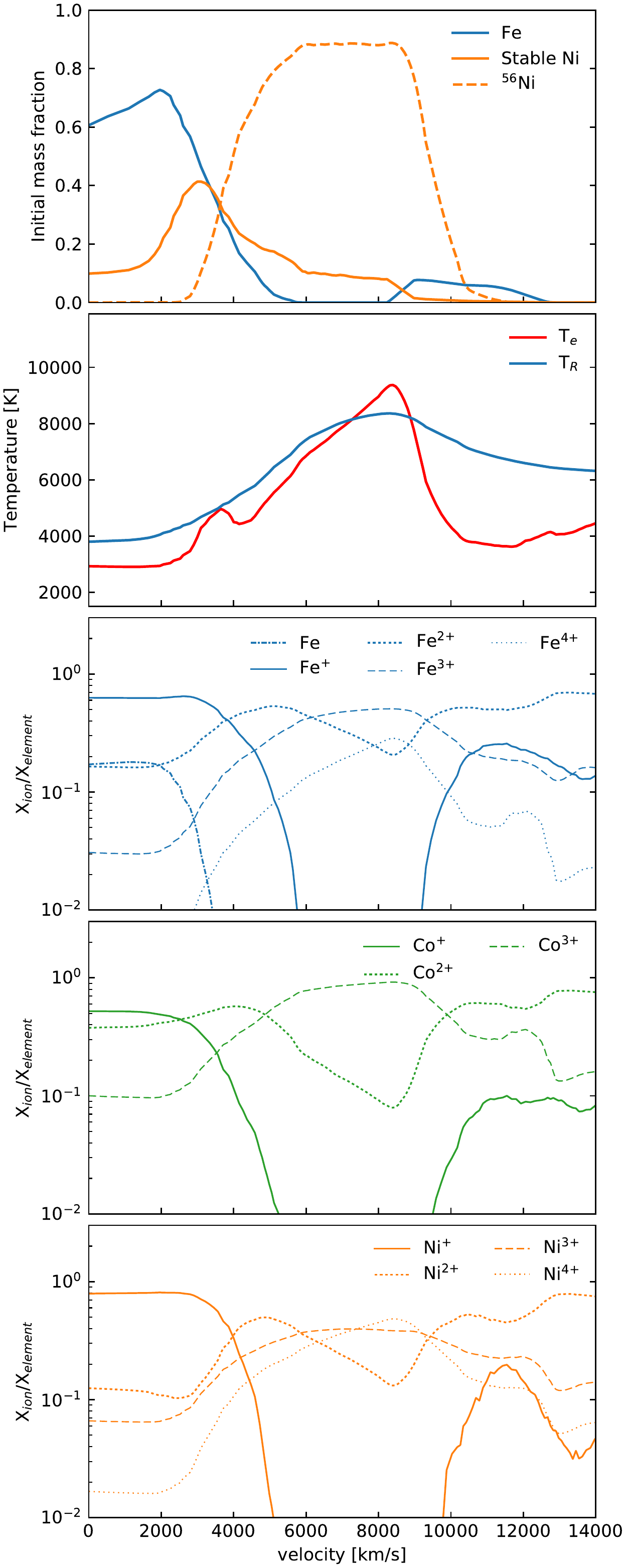}\end{center}
 \caption{Composition immediately after explosion (top panel), and at 330 days: the electron temperature ($T_\mathrm{e}$) and radiation temperature ($T_\mathrm{R}$) (second panel), Fe ion balance (third panel), Co io balance (fourth panel), and Ni ion balance (bottom panel) as a function of velocity for the inner ejecta of the \artis W7 model. By this time, most of the initial \iso{56}{Ni} will have decayed into \iso{56}{Co} and \iso{56}{Fe}.
\label{fig:330d-w7-estimators}}
\end{figure}

The structure of the inner ejecta is of primary relevance to understanding late-phase supernova spectra (the outer ejecta having expanded sufficiently to be optically thin in most places).
The inner part of the W7 model includes a core of stable Ni isotopes and Fe (produced by electron capture during the explosion), which is surrounded by a \iso{56}{Ni}-rich region.
This composition profile is shown in the top panel of \autoref{fig:330d-w7-estimators}.

In our calculation (epoch of 330 days post explosion), we find that the core of stable material is at low temperature ($\sim 3000$~K) and is mostly singly-ionised (temperature and ionisation fractions are shown in \autoref{fig:330d-w7-estimators}).
The temperature and ionisation then rise rapidly through the (initially) \iso{56}{Ni}-rich region leading to a peak of $\sim 9000$~K and predominantly two-to-three times ionised material in the ejecta around $\sim 8000$~km~s$^{-1}$.
This thermal/ionisation structure is in generally good (albeit imperfect) agreement to that calculated for the W7 model by \cite{Liu:1997hs} (see their figure 1, which shows ejecta properties for an epoch of 300~days).
One difference is that our Fe$^{{+}}$ abundance fraction becomes very low (below 10$^{-2}$) in the \iso{56}{Ni}-rich region between 6000 to 9000 kms$^{-1}$, due to the high photoionisation rate in these zones.

\subsubsection{Non-LTE populations}
\label{sec:nlte-pops}
\begin{figure*}
 \begin{center}\includegraphics[width=\textwidth]{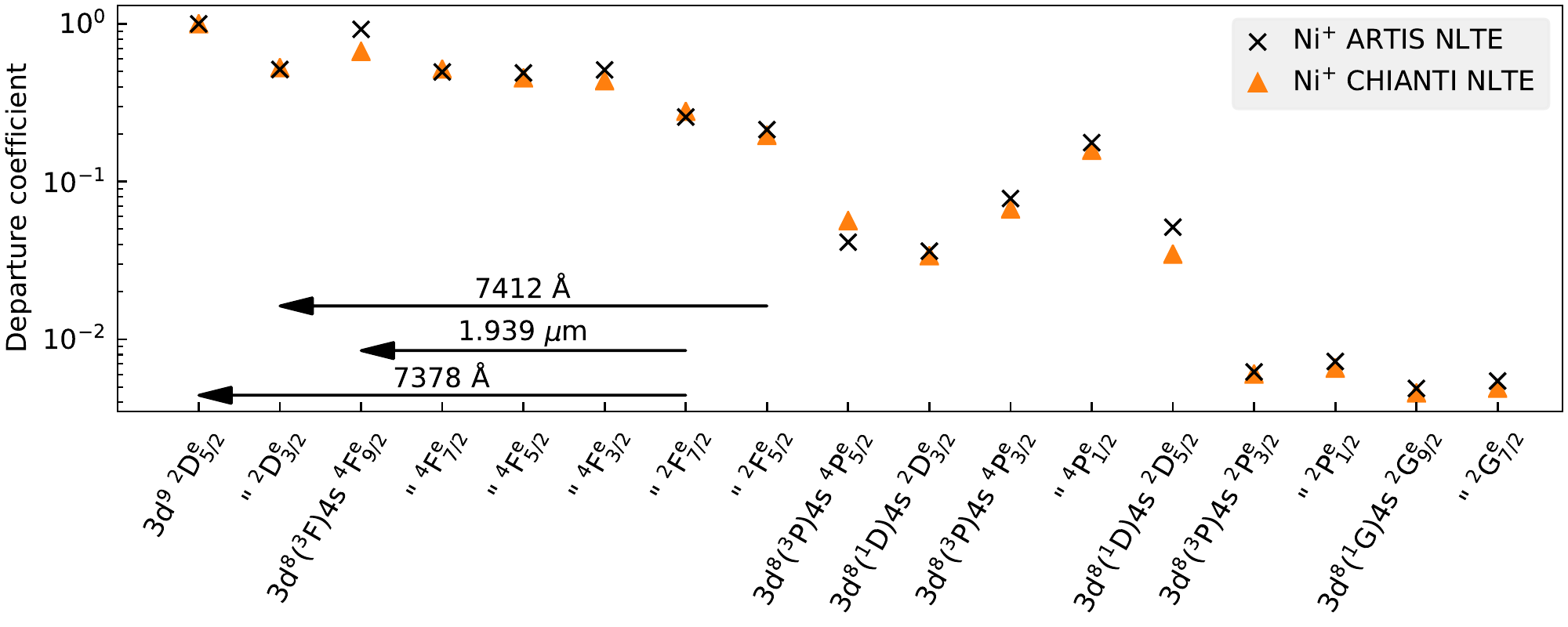}\end{center}
 \caption{Non-LTE departure coefficients (population relative to the ground state, divided by the same quantity for Boltzmann equilibrium at the electron temperature) for \ion{Ni}{ii} in a \iso{56}{Ni}-rich region near 4000 km/s in the W7 model at 330 days.
 Also shown is reference data from CHIANTI in a collisional-radiative model at the same electron temperature (7940.1 K) and electron density ($1.1 \times 10^6$ cm$^{-3}$).
 The arrows represent several important transitions that are the most likely to produce observable spectral features - the $\lambda\lambda$7378, 7412 doublet and the 1.939 $\mu$m near-IR line.
 \label{fig:w7nlte-ni}}
\end{figure*}

A key improvement in the new version of \artis used here is the implementation of our NLTE population solver.
As one example of the importance of the strength of non-LTE effects,
\autoref{fig:w7nlte-ni} shows the calculated departure coefficients of the first 17 levels of \ion{Ni}{ii} in one of the \iso{58}{Ni}-rich model cells (selected at 4000 km~s$^{-1}$) of our W7 calculation (at 330 days post explosion).
This demonstrates that the populations of excited states that are responsible for observable emission features in SNe~Ia spectra can be expected to depart significantly from LTE \citep{Axelrod:1980vk}.

As a test of our NLTE populations, we compared our departure coefficients to those calculated with a collisional-radiative model, which assumes statistical equilibrium between thermal electron collisions and radiative decays.
Specifically, we utilised the CHIANTI atomic database and analysis package \citep{Dere:1997gd,DelZanna:2015ie} to compute departure coefficients adopting the electron temperature and free electron density obtained from \artis.
Although this is a substantial simplification compared to the full treatment in the code, the relatively high electron temperature ($T_e = 7940.1$~K) and low radiation field intensity obtained in this region of the model at this time means that the populations of the first few levels are primarily controlled by thermal electron collisions and spontaneous radiative decay. The CHIANTI model includes only 17 levels, which causes the departure coefficients to diverge from our model from around level 9 and above. These levels are significantly populated by radiative decays from higher levels that are not included in CHIANTI.
As shown in \autoref{fig:w7nlte-ni}, the level of agreement is generally good (typically a few tens per cent, or better).

\subsubsection{Radiation field consistency check}

\begin{figure}
 \begin{center}\includegraphics[width=\columnwidth]{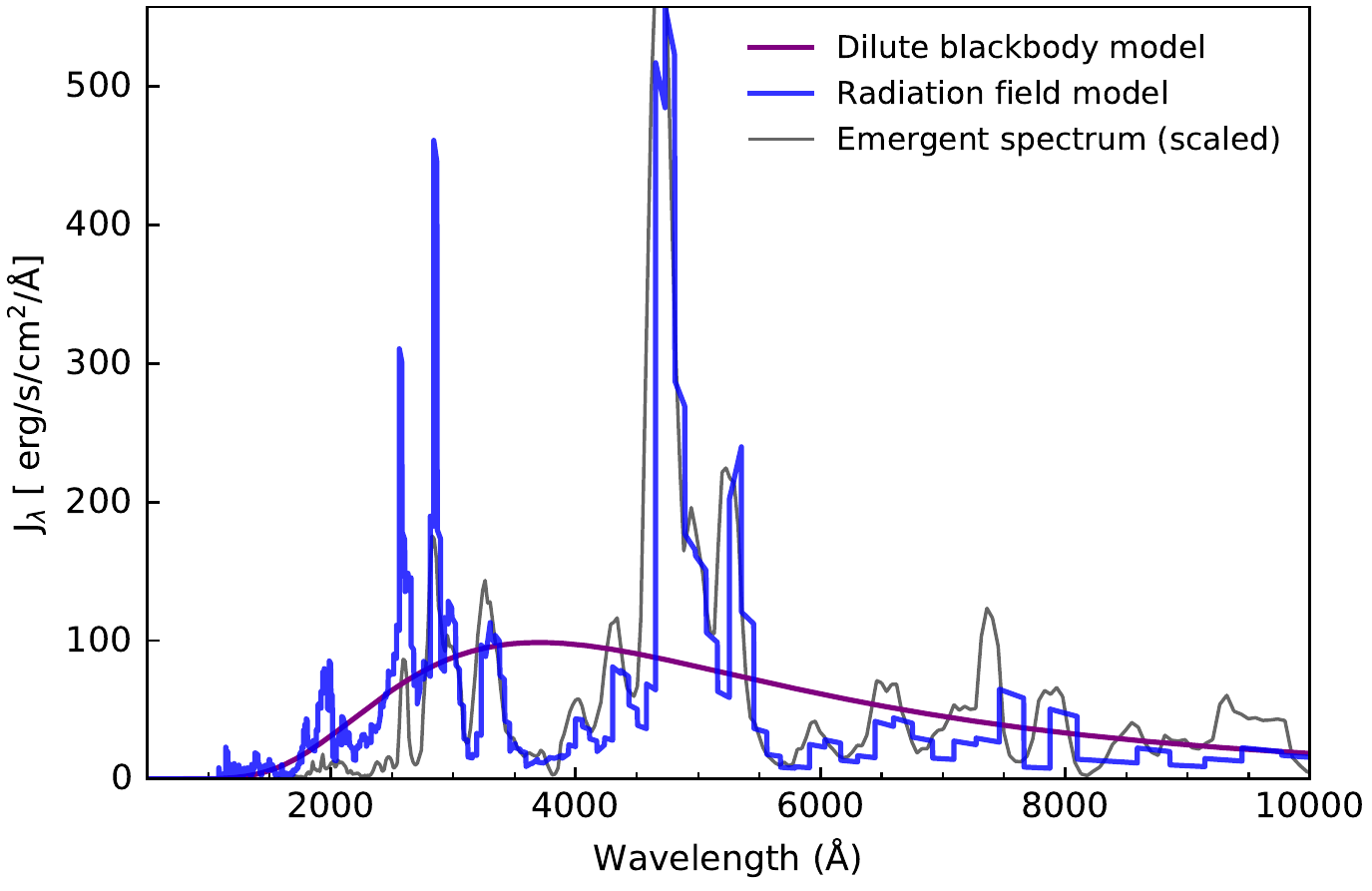}\end{center}
 \caption{Radiation field estimators compared with the emergent spectrum for an (initially) \iso{56}{Ni}-rich cell near 7000 km/s in the W7 model at 330 days. This figure shows the full-spectrum fit dilute blackbody field (purple), and the binned dilute blackbody fits (blue) which describe the radiation field for \artis in this work. Because this cell has a relatively high radiation temperature, the radiation field is typically greater at bluer wavelengths compared to the outgoing spectrum. \label{fig:330d-w7-radfield}}
\end{figure}

\autoref{fig:330d-w7-radfield} shows the internal radiation field reconstructed from our Monte Carlo estimators (see Section \ref{sec:radiationfield}) for an (initially) \iso{56}{Ni}-rich zone at 7000 km/s in the W7 model at 330 days.
For comparison, \autoref{fig:330d-w7-radfield} also shows a single dilute blackbody model for the full-spectrum and the external radiation field (i.e.\ emergent simulation spectrum) scaled to the radius of the cell.
The internal radiation field has a similar spectral energy distribution to the emergent radiation field (aside from a Doppler shift, see \autoref{fig:w7comparison}), as would be expected for the approximately optically-thin conditions of the nebular phase.
Clearly, the radiation field model represents an improvement compared to the full-spectrum dilute blackbody model in previous \artis simulations.

\subsubsection{W7 nebular spectrum comparison}
\label{sec:w7comparison}

In \autoref{fig:w7emissionspectra} we show the computed spectrum in the optical and near-infrared (near-IR) regions for our \artis W7 calculation at 330 days with colour coding to illustrate the ions responsible for the emission.
Specifically, each of our Monte Carlo quanta is tagged by their last  interaction with the thermal pool (i.e. their last $k$-packet interaction in the nomenclature of \citealt{Lucy:2002hw}).
This gives an indication of the contributions of the different ions to the cooling of the ejecta.
Note, however, that some packets do undergo fluorescence/scattering by other ions prior to escape from the simulation, which is not captured by this simple tagging scheme.

In agreement with previous studies of nebular spectra for SNe~Ia \citep{Axelrod:1980vk,RuizLapuente:1992iu,Liu:1997hs}
the strongest peak of the spectrum is formed mostly by [\ion{Fe}{iii}] $\lambda\lambda$4658, 4701 emission, with contributions from other transitions between the states in the low-lying $^3$F2 term and $^5$D ground term.
Much of the rest of the spectrum in the optical is due to [\ion{Ni}{ii}] and [\ion{Ni}{iii}], and the near-IR is dominated by [\ion{Fe}{ii}], with several strong contributions from [\ion{Ni}{ii}].
Strong features of [\ion{Ni}{ii}] in particular are the
$\lambda\lambda$7378, 7412 doublet (3p$^6$3d$^8$($^3$F)4s $^2$F$_{7/2}$ $\rightarrow$ 3p$^6$3d$^9$ $^2$D$_{5/2}$ and 3p$^6$3d$^8$($^3$F)4s $^2$F$_{5/2}$ $\rightarrow$ 3p$^6$3d$^9$ $^2$D$_{3/2}$) in the optical and [\ion{Ni}{ii}] (3d$^8$($^3$F)4s $^2$F$_{7/2}$ $\rightarrow$ 3d$^8$($^3$F)4s $^4$F$_{9/2}$) at 1.939 $\mu$m in the near IR.
As has been previously noted, the W7 model predicts relatively strong [\ion{Ni}{ii}] compared to observations \citep[see e.g.,][]{Liu:1997hs}, and we also find similarly strong features to \citet{Liu:1997hs}, \citet{RuizLapuente:1995cv}, \citet{Sollerman:2004fy}, \citet{Maeda:2010fx}, \citet{Maurer:2011cz}, \citet{Mazzali:2011es}, and \citet{Fransson:2015ct}.

The lateness of the epoch considered here means that Co emission is comparatively weak, and we do not find significant influence from other elements across most of the spectral region considered: the most notable contribution is from \ion{S}{iii} between 9000 and 10000~\AA.
This feature is due to emission from the [\ion{S}{iii}] $\lambda\lambda$9069, 9530 lines (3s$^2$3p$^2$ $^1$D$_2$ $\rightarrow$ $^3$P$_{1,2}$).

\begin{figure}
 \begin{center}\includegraphics[width=\columnwidth]{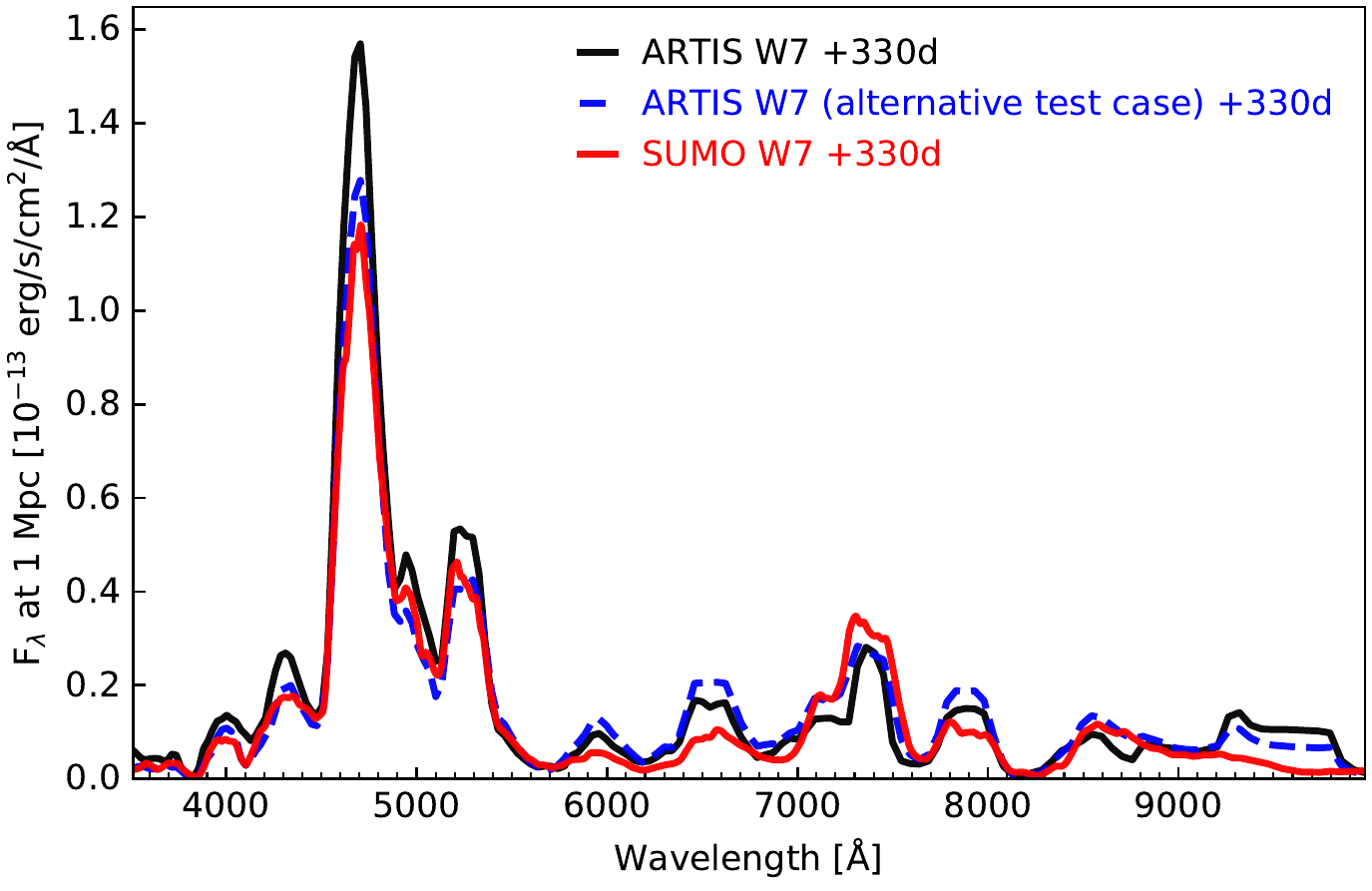}\end{center}
 \caption{Nebular spectra of the W7 model at 330 days calculated with \artis (black) and with \sumo (red). We also show (in blue) the spectrum from a test \artis calculation (see text). \label{fig:w7comparison}}
\end{figure}

\begin{figure*}
 \begin{center}\includegraphics[width=0.5\textwidth]{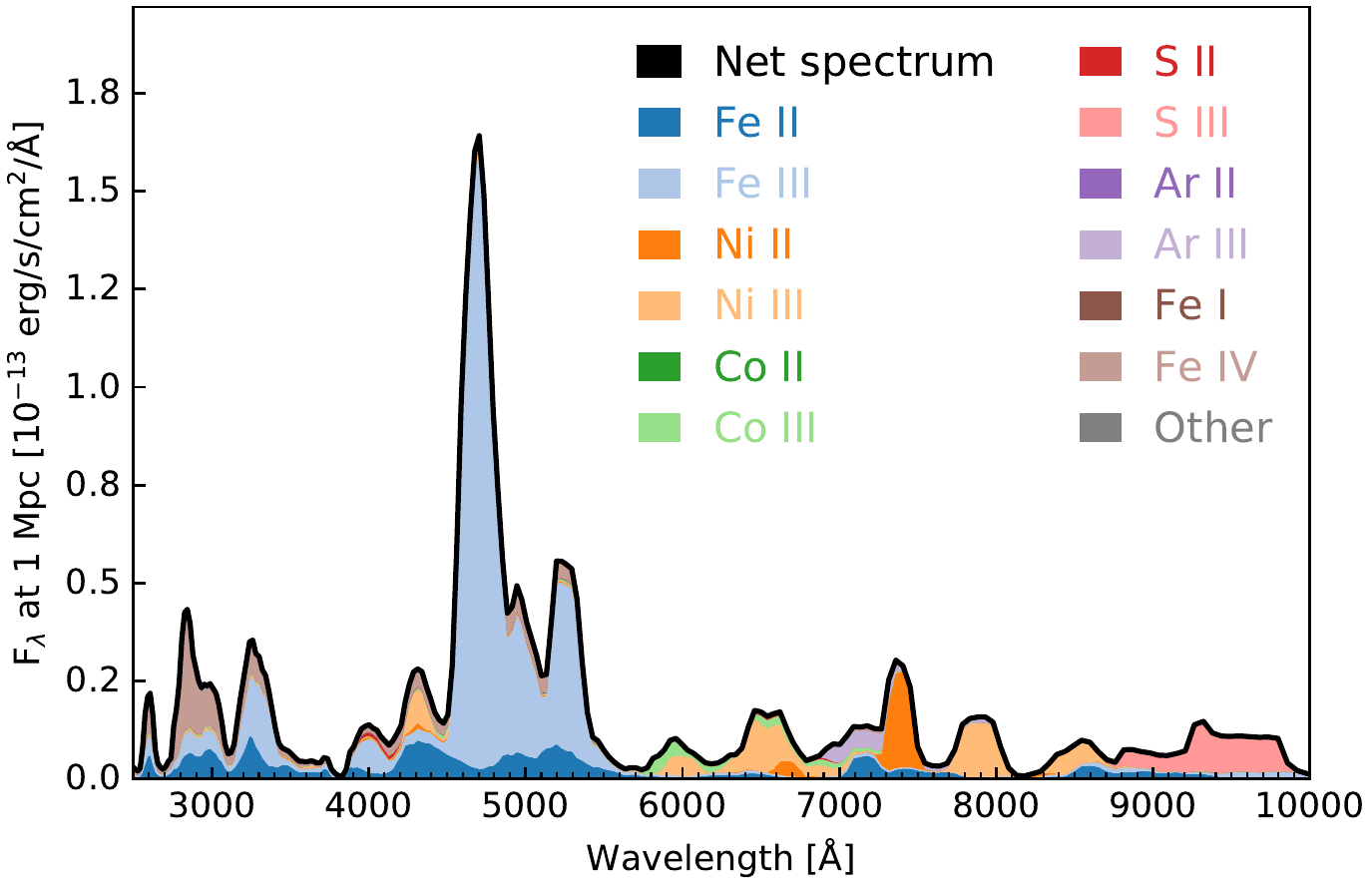}\includegraphics[width=0.5\textwidth]{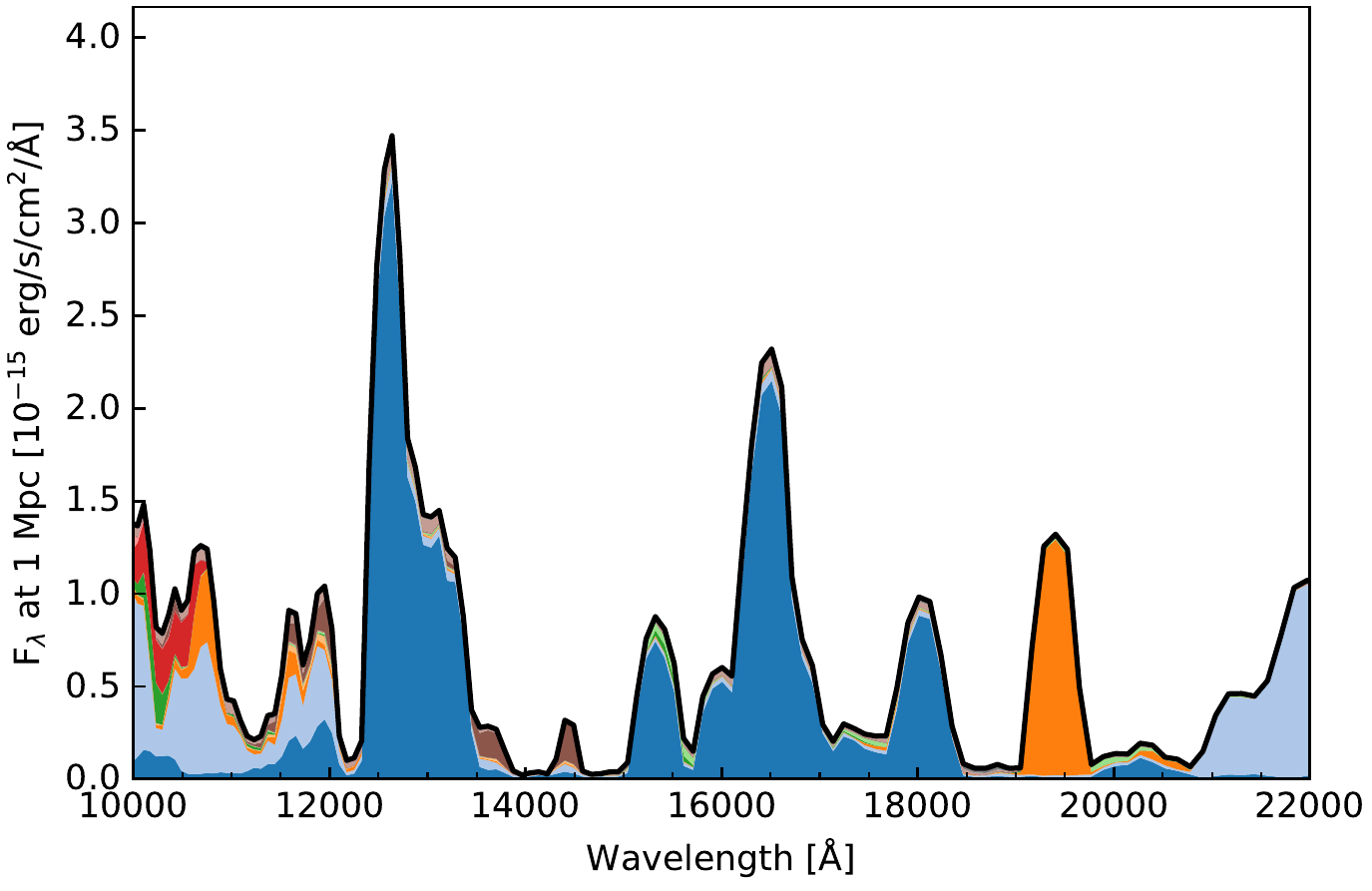}\end{center}
 \caption{Nebular emission spectra of the W7 model at 330 days in the optical (left) and near-infrared (right). The total spectrum is plotted as the black curve. The area under the spectrum is colour coded to indicate which ions are responsible for the emission in each wavelength bin (see text). Note the differing scales on left and right panels.
\label{fig:w7emissionspectra}}
\end{figure*}

In \autoref{fig:w7comparison} we compare our optical spectrum for the W7 model at 330 days after explosion with the spectrum of \citet{Fransson:2015ct}\footnote{Available at \url{https://star.pst.qub.ac.uk/webdav/public/ajerkstrand/Models/FranssonJerkstrand2015/W7_330d.dat}}, which has been calculated using the \sumo radiative transfer code described by \citet{Jerkstrand:2011fz} with updates described by \citet{Jerkstrand:2015dz}.
We note that the \sumo spectrum of W7 at 330 days is generally similar to the spectra produced by the \citet{Mazzali:2001gz} and \textsc{nero} codes \citep[figures 5 and 6 of][]{Maurer:2011cz}.
In general, the spectra from the \artis and \sumo calculations are extremely similar.
However, there are several minor quantitative differences which may be related to differences in approach.
%Firstly, we note that the \artis calculation predicts weaker emission in the region dominated by [\ion{Fe}{iii}] but stronger emission by [\ion{Fe}{ii}] (particularly in the near-UV) and [\ion{Ni}{ii}].
%The [\ion{Fe}{ii}] emission bump around 4300~\AA~is also very noticeably different in structure.

%Given the complexity of the calculations, it is challenging to fully identify the origins of all discrepancies.
%Nevertheless, it is likely that some of the differences found here can be directly attributed to simple differences in the adopted atomic data and treatment of atomic processes.

The \sumo calculation includes neutral species and ions up to doubly-ionised for Fe, Co and Ni,
but does not consider higher ions (i.e. Fe$^{3+}$ and above). In contrast, our calculation neglects neutral Co and Ni but includes the triply-ionised species for all three of these elements.
\sumo includes several hundred non-LTE levels for Fe-group elements, while we use 80 for most ions, and 197 for Fe$^{{+}}$.
Another difference is that the \sumo calculation substitutes the valence-shell potential for the inner-shell potential when calculating the ionisation rates (Equation~\ref{eqn:ionisationrate}) as a way of compensating for multiple ionisations resulting from Auger electrons.
In \artis, we use the individual shell potentials and their Auger yields (see Section~\ref{sec:ntrates}) to determine the rates of single, double, and triple ionisation. It may therefore be expected that our approach may provide more accurate ionisation rates, while the \sumo treatment may give an upper bound on the total ionisation rates. However, our testing found that the consequences of these multiple ionisations for the resulting spectra are a minor effect.

To explore some of these issues, we have carried out a second \artis test calculation (blue dashed line in \autoref{fig:w7comparison}) in which we (i) do not include Fe$^{3+}$, Co$^{3+}$ or Ni$^{3+}$ (or higher ions), (ii) divide ionisation energies by the valence potentials and set the Auger probabilities to zero, and (iii) use the \cite{Shull:1982ib} recombination rate for Ni$^{2+}$.
This test calculation fits the [\ion{Fe}{ii}] and [\ion{Fe}{iii}] features of \sumo very precisely.
However, the [\ion{Ni}{iii}] features (which were already stronger than \sumo in our reference calculation) become even stronger, which is probably mostly due to the slower recombination rate of Ni$^{2+}$ to Ni$^{{+}}$.

The \sumo calculation does not include [\ion{S}{iii}] emission lines and thus does not match \artis in the region around $\sim 9000 - 10000$~\AA.
We note, however, that a similar feature is found in the nebular spectra calculated by \citet{Wilk:2018ki}.

\begin{figure*}
 \begin{center}\includegraphics[width=0.5\textwidth]{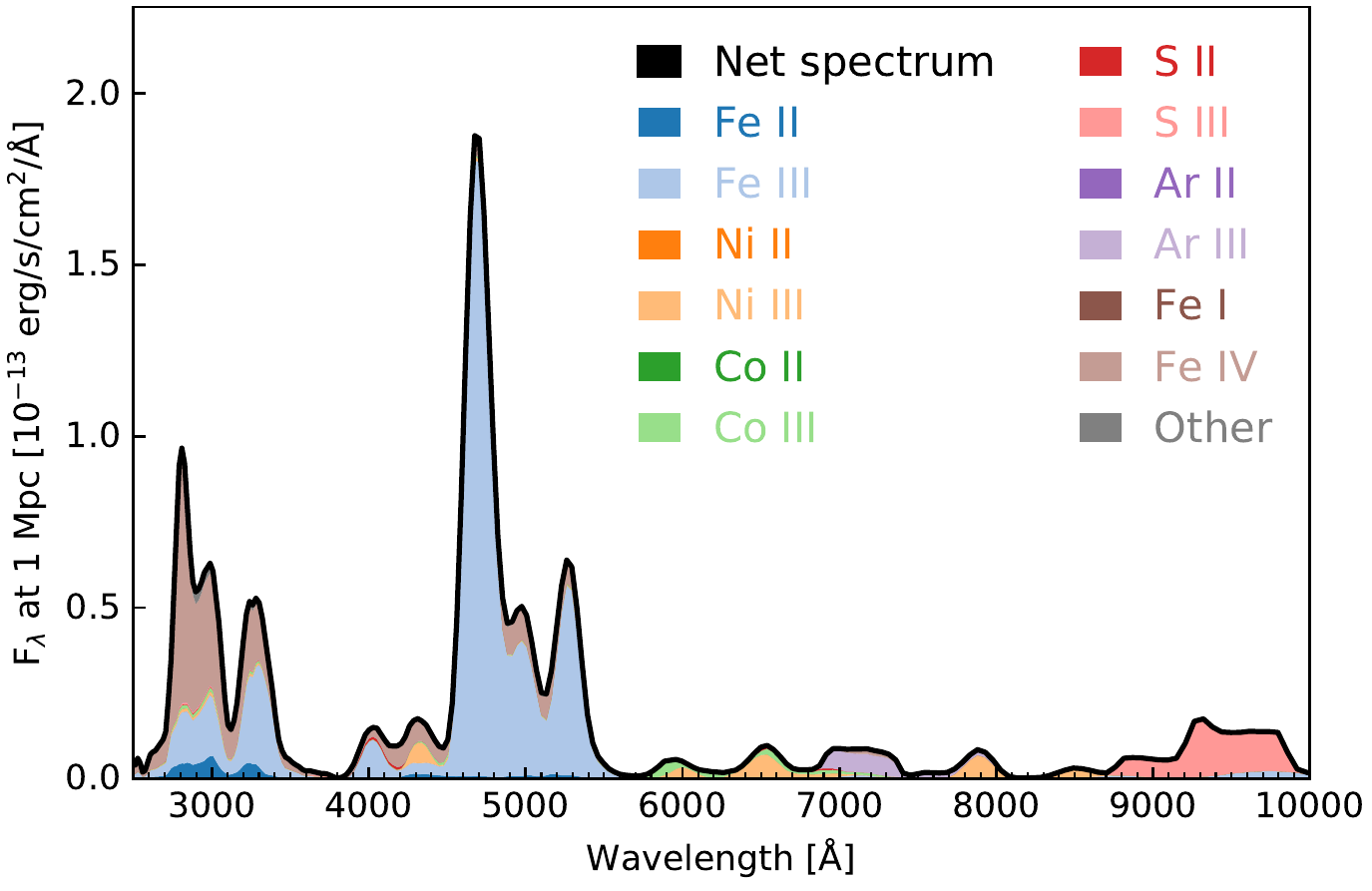}\includegraphics[width=0.5\textwidth]{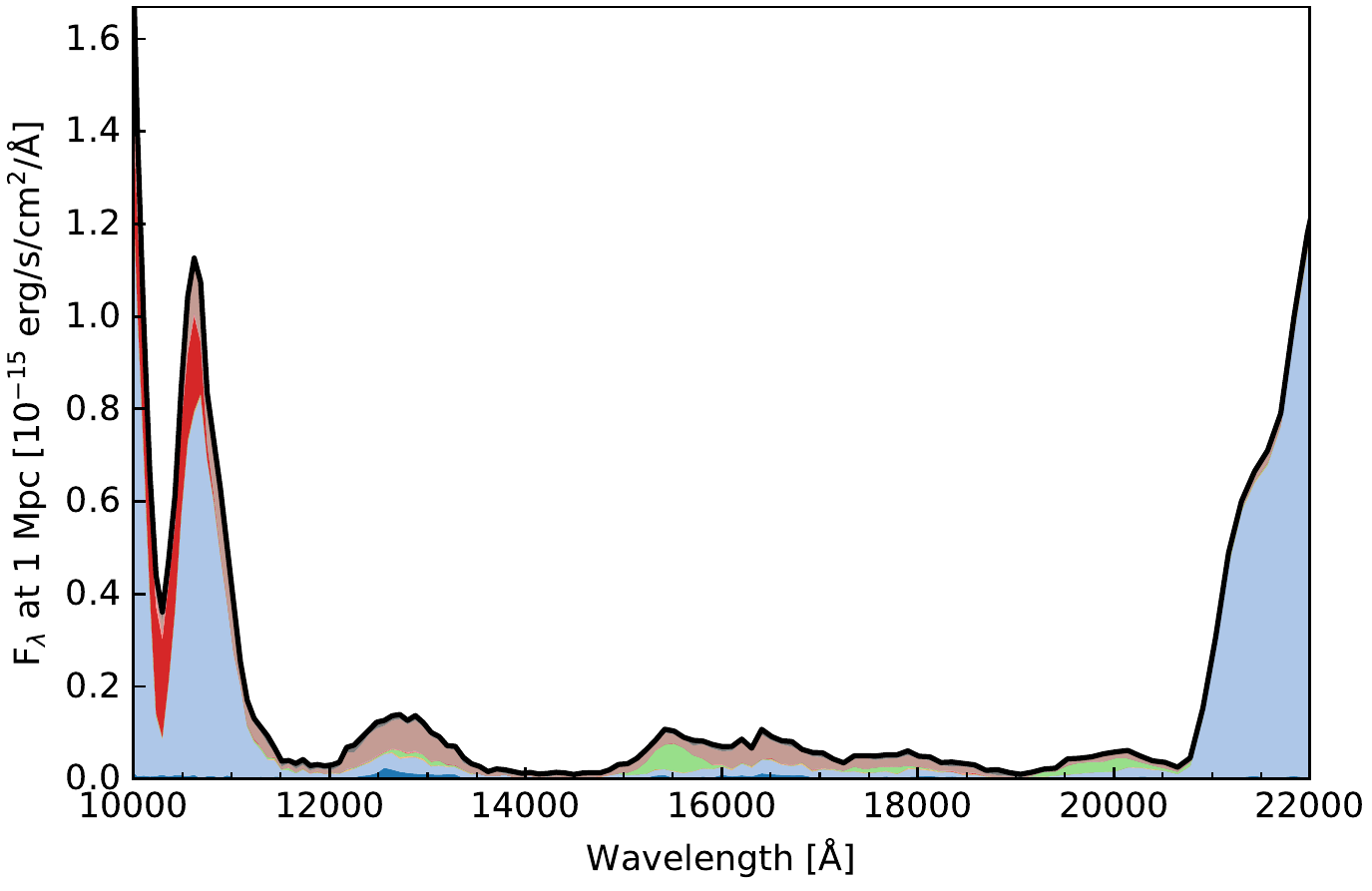}\end{center}
  \caption{Nebular spectra coloured by ion for S0 \artis models at 330 days in the optical (left panels) and near-infrared (right panels). Note the differing scales on left and right panels.
\label{fig:emissionspectra_s0}}
\end{figure*}

\begin{figure*}
 \begin{center}\includegraphics[width=0.5\textwidth]{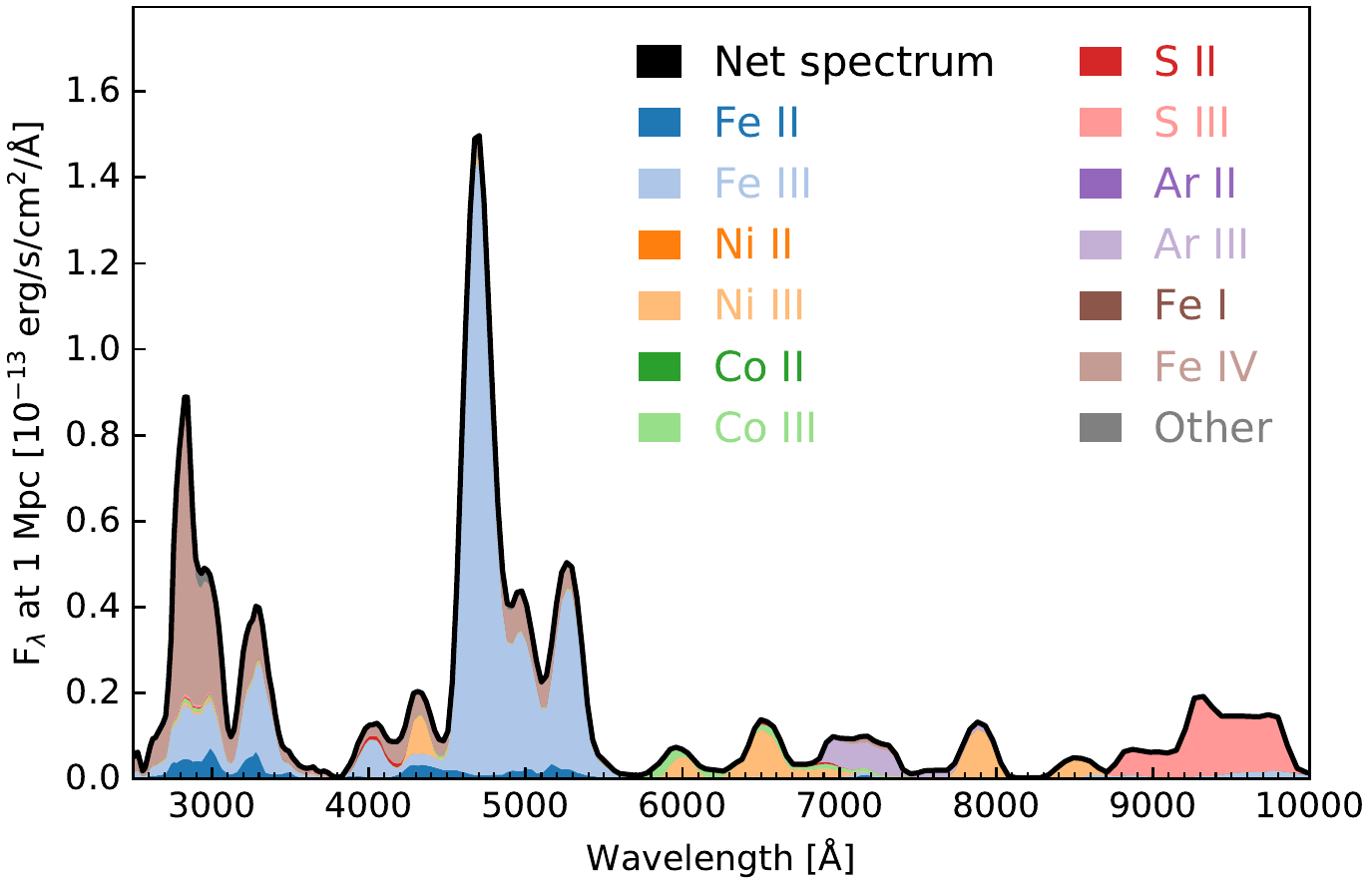}\includegraphics[width=0.5\textwidth]{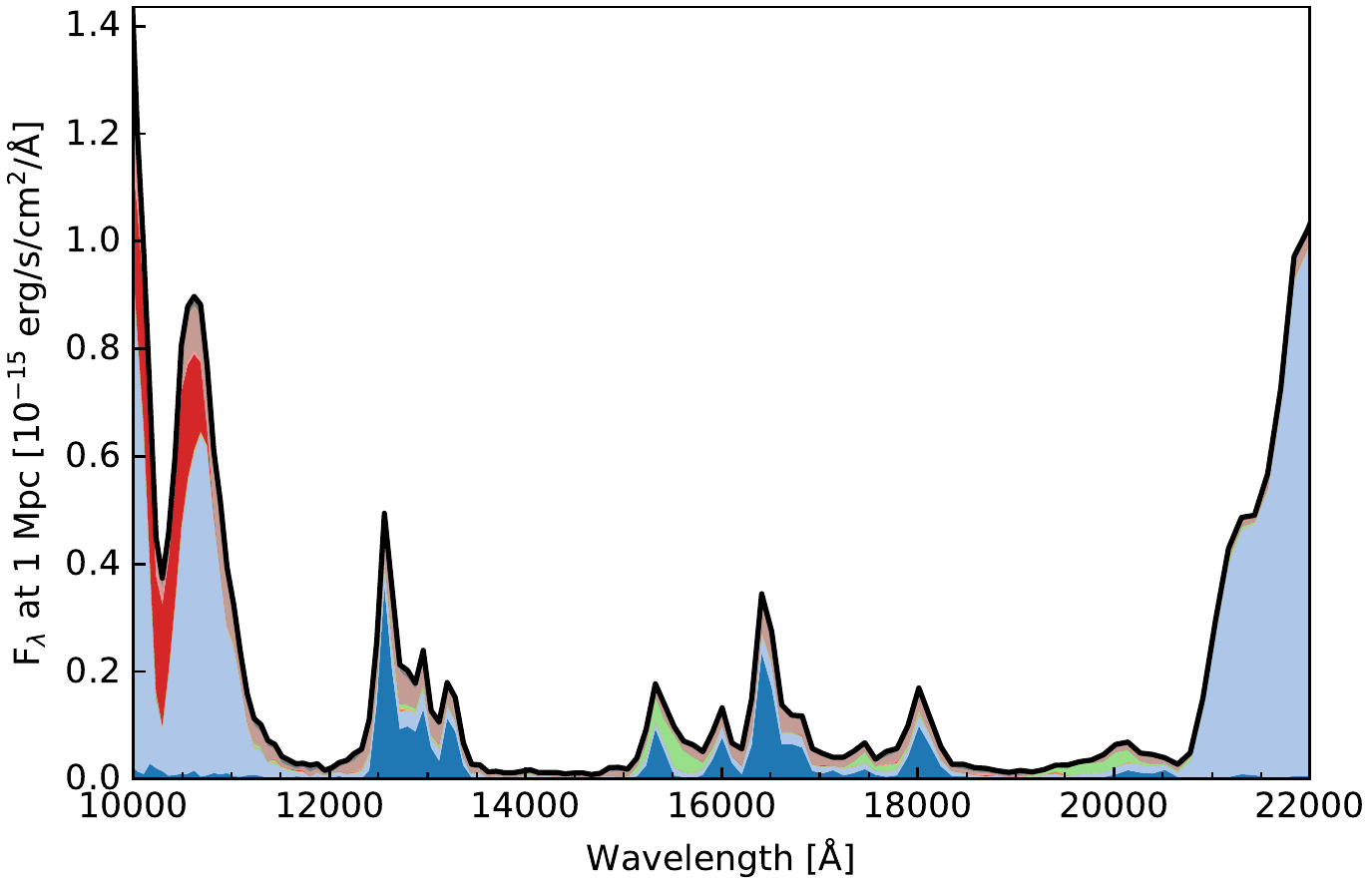}\end{center}
 \caption{Nebular spectra coloured by ion for S5.5 \artis models at 330 days in the optical (left panels) and near-infrared (right panels).
Note the differing scales on left and right panels.
\label{fig:emissionspectra_s5p5}}
\end{figure*}

\section{Sub-Chandrasekhar detonation models}
\label{sec:subchdet}

As a first application of the code developments described above, we present nebular-phase spectra for sub-\Mch detonation models.

\subsection{Motivation}

The relatively low densities in sub-\Mch explosion models makes electron capture inefficient when compared to \Mch explosion scenarios.
Consequently the yields of neutron-rich isotopes, such as \iso{54}{Fe} and \iso{58}{Ni}, are relatively small compared to Chandrasekhar mass deflagration models.
This is one promising way to distinguish Chandrasekhar mass and sub-\Mch explosion models.
In particular, \iso{58}{Ni} is of special interest since its presence is directly probed by nebular phase spectra (although \iso{58}{Ni} is a minor contribution to the overall Ni abundance immediately post-explosion, the relatively rapid decay of \iso{56}{Ni} means that \iso{58}{Ni} becomes dominant at late times).

However, sub-\Mch models do produce some stable Fe-group material owing to the presence of the neutron-rich nuclide \iso{22}{Ne} in the progenitor.
This isotope increases the neutron-to-proton ratio in the fuel, which when burned to nuclear statistical equilibrium, results in an enhanced production of stable Fe-group nuclides \citep[e.g.,][]{Timmes:2003bp,Seitenzahl:2017gu}.
Understanding the influence of \iso{22}{Ne} in sub-\Mch models is potentially complicated by the action of gravitational settling in WD stars.
This process can cause the neutron-rich \iso{22}{Ne} nuclei to accumulate near the centre \citep{Deloye:2002gs,GarciaBerro:2008gj}.
As a consequence, the explosion of a gravitationally-settled WD model may produce more stable Fe-group material near the centre. This is somewhat reminiscent of scenarios in which the inner ejecta have a high concentration of stable Fe-group elements as found in 1D delayed-detonation models \citep[and has been inferred from observations, e.g.,][]{Mazzali:2007ix}. However, the effect is different in origin and much less pronounced (i.e. although \iso{22}{Ne} settling may lead to an enhanced abundance of stable material in the inner ejecta, it is not expected that it will produce a core dominated by stable isotopes; see also \citealt{Bravo:2011fa}).
Here, our objective is to quantify this issue by calculating nebular phase spectra for simple detonation models of sub-\Mch WDs in which the progenitor WD contains different radial profiles of \iso{22}{Ne}.

\subsection{Adopted models}

We study two specific sub-\Mch detonation models.
They were calculated by \citet{Michel:2014uw} using the hydrodynamics code \textsc{leafs} \citep{Reinecke:2002ij, Ropke:2005cu} following the same approach described by \cite{Sim:2010hl} and \cite{Marquardt:2015hm}.

In both cases, the initial WD is hydrostatic and has a total mass of 1.06\,M$_{\odot}$.
The models differ, however, in the adopted composition structure of the WD: specifically in the distribution of \iso{22}{Ne}.
In the first model (hereafter `S0' model), a uniform mass fraction of $X(^{22}{\rm Ne})=0.02$ (corresponding to approximately solar metallicity) is adopted throughout the WD: i.e.\ this represents the limit of no settling.

\begin{figure}
 \begin{center}\includegraphics[width=0.5\textwidth]{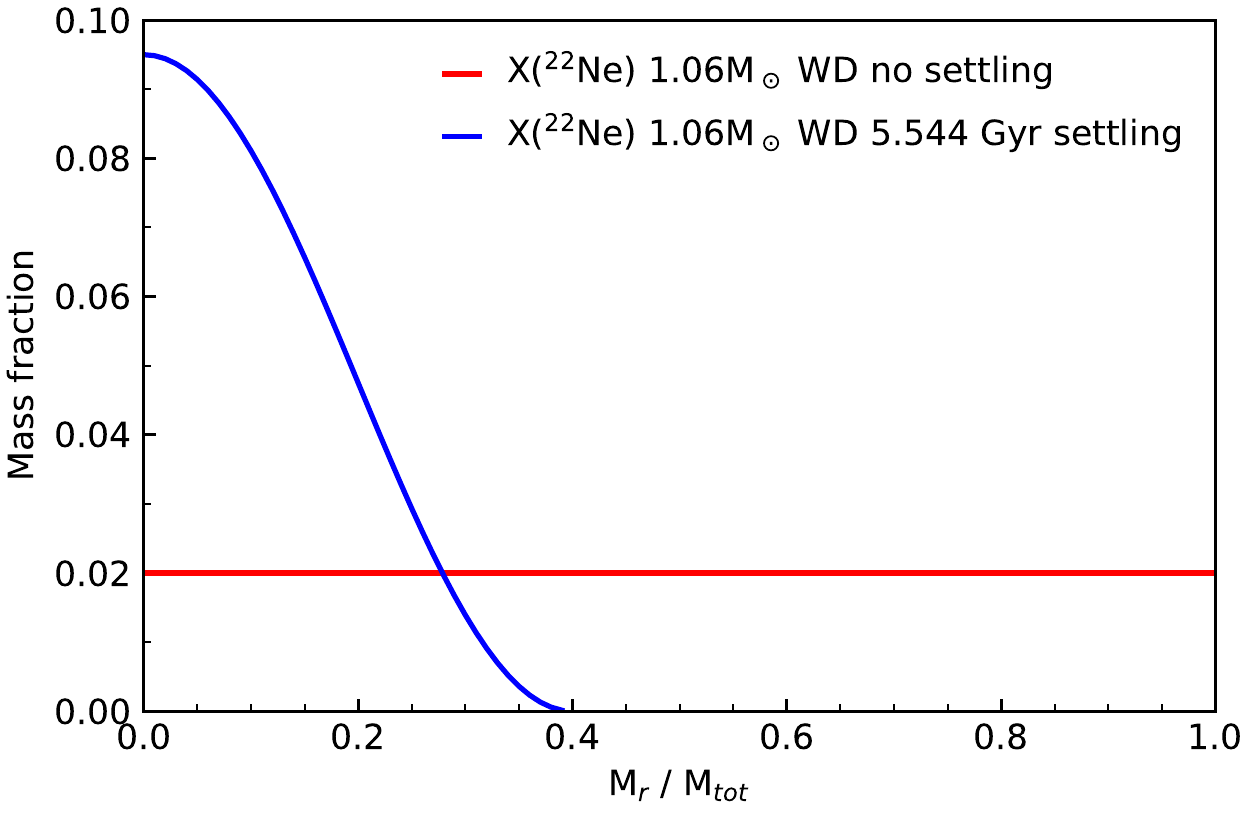}\end{center}
 \caption{\iso{22}{Ne} profiles adopted for the WD pre-explosion models with no settling (S0, red) and after 5.544 Gyr of boosted gravitational settling (S5.5, blue).\label{fig:settlingprofiles}}
\end{figure}

In the second model, `S5.5', a $X(\iso{22}{Ne})$ profile was constructed to approximately correspond to an extremely high degree of \iso{22}{Ne} settling.
In particular, this is based on the most extreme settling calculated for a 1.06\,M$_{\odot}$~WD in the study by \cite{GarciaBerro:2008gj}
(settling time of 5.544\,Gyr and a boosted diffusion coefficient $D = 5 D_S$ where $D_S$ is the diffusion coefficient due to gravitational settling; see
equation (18) of \citealt{GarciaBerro:2008gj} for details).
For the specific simulation carried out here, $X(\iso{22}{Ne})$ has a maximum value of 0.095 at the centre of the WD and decreases monotonically outwards to $X(\iso{22}{Ne})= 0$ at mass coordinate $M = 0.4$~M$_{\odot}$ (for simplicity, a sine-squared functional form was adopted, see \autoref{fig:settlingprofiles}).
In both models it was assumed that the WD initially consists of C and O in equal parts.
We further assume that C and O are substituted in equal parts during the settling of \iso{22}{Ne}.
This is a simplification since the settling depends on the proton number $Z$ and the nucleon number $A$ of the surrounding material and also differs from \cite{GarciaBerro:2008gj} who studied the \iso{22}{Ne} settling in a pure C environment for simplicity.

The hydro-dynamical explosion simulations were performed in the manner described by \cite{Marquardt:2015hm}, including calibration of detonation energy-release tables for the appropriate compositions.
Sets of nucleosynthesis tracer particles were then processed using the 384-isotope network of \cite{Travaglio:2004bp}, from which the ejecta profiles were derived for the homologous expansion phase (see \citealt{Marquardt:2015hm} for a description of the methods).

As expected from the discussion above, the most important difference in structure between the two explosion models is in the distribution of Fe-group elements.
In particular, the \iso{56}{Ni} mass fraction in the S0 model is almost uniform throughout the inner $\sim 8000$~km~s$^{-1}$ of the model (accompanied by an approximately uniform composition of stable Fe and Ni).
In contrast, \iso{22}{Ne} settling in the S5.5 model leads to an enhanced concentration of stable Ni in the inner core with a maximum in the \iso{56}{Ni} distribution around $8000$~km~s$^{-1}$ (see \autoref{fig:330d-subch-estimators}).
This central concentration of stable isotopes is qualitatively reminiscent of the W7 model (see \autoref{fig:330d-w7-estimators}) although {\it much} less pronounced.
The enhanced production of \iso{56}{Ni} at the expense of stable Fe outside 8000~km~s$^{-1}$ in the S5.5 model is due to the lack of \iso{22}{Ne} in the outer layers.

\begin{table}
{
\centering
	\caption{Synthesised masses of \iso{56}{Ni}, \iso{54}{Fe}, and \iso{58}{Ni} from the W7, S0, and S5.5 explosion models.}
	\label{tab:explosionmodels}
	\begin{tabular}{lccr}
		\hline
		& \multicolumn{3}{c}{Synthesised mass}\\
		Model & \iso{56}{Ni} & \iso{54}{Fe} & \iso{58}{Ni}\\
		 & $[\Msun]$ & $[10^{-2}\,\Msun]$ & $[10^{-2}\,\Msun]$\\
		\hline
		W7$^a$	& 0.59 & 9.5 & 11.0\\
		S0		& 0.56 & 2.0 & 1.8\\
		S5.5	& 0.55 & 1.2 & 3.4\\
		\hline
	\end{tabular}\\}
    $^a$The W7 nucleosynthesis used here is that of \cite{Iwamoto:1999jd}. We note that significantly lower yields of \iso{58}{Ni} have been found in the updated calculations described by \cite{Nomoto:2018iw}.
\end{table}

\autoref{tab:explosionmodels} shows the synthesised masses of \iso{56}{Ni}, \iso{54}{Fe}, and \iso{58}{Ni} from the W7, S0, and S5.5 explosion models.
Compared to the S0 model, the S5.5 model has synthesised a few per cent less \iso{56}{Ni}, 40 per cent less \iso{54}{Fe}, and of particular importance for the nebular spectra, 85 per cent more \iso{58}{Ni} mass.

\subsection{Overview of calculations}

\begin{figure*}
 \begin{center}\includegraphics[width=0.5\textwidth]{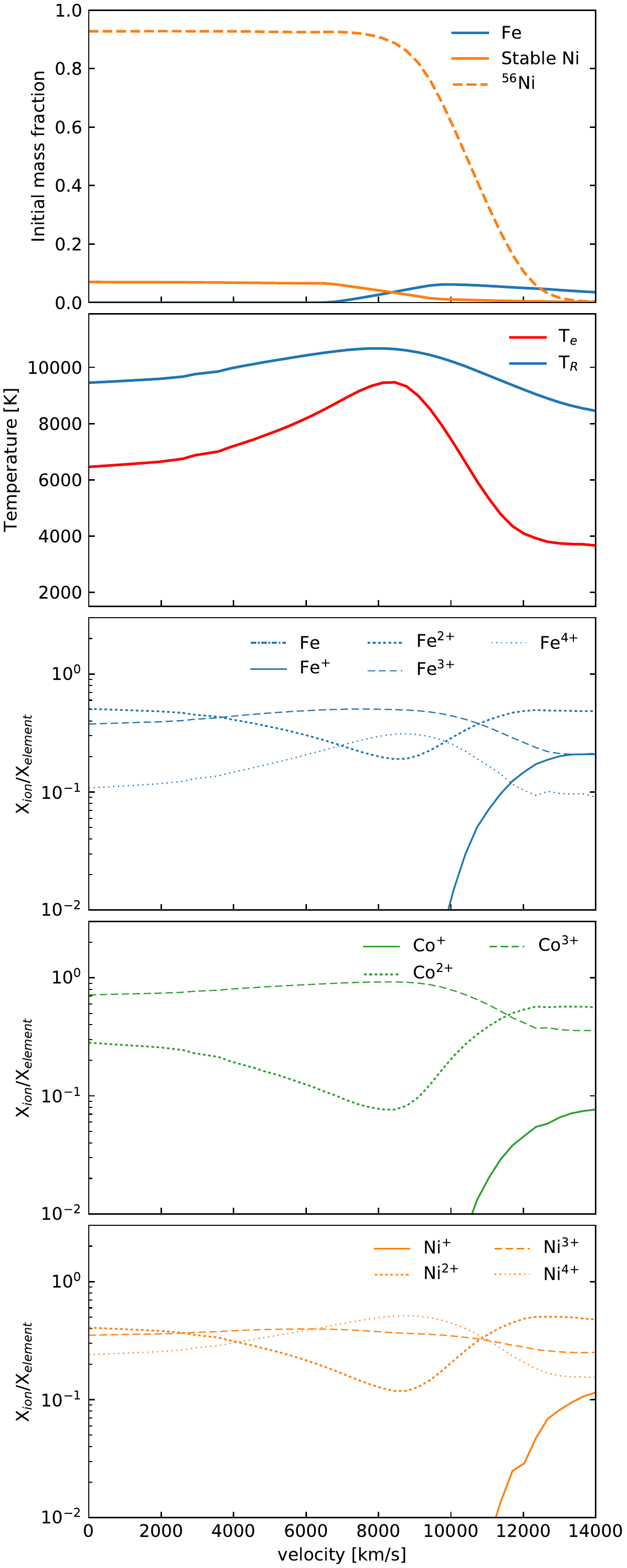}\includegraphics[width=0.5\textwidth]{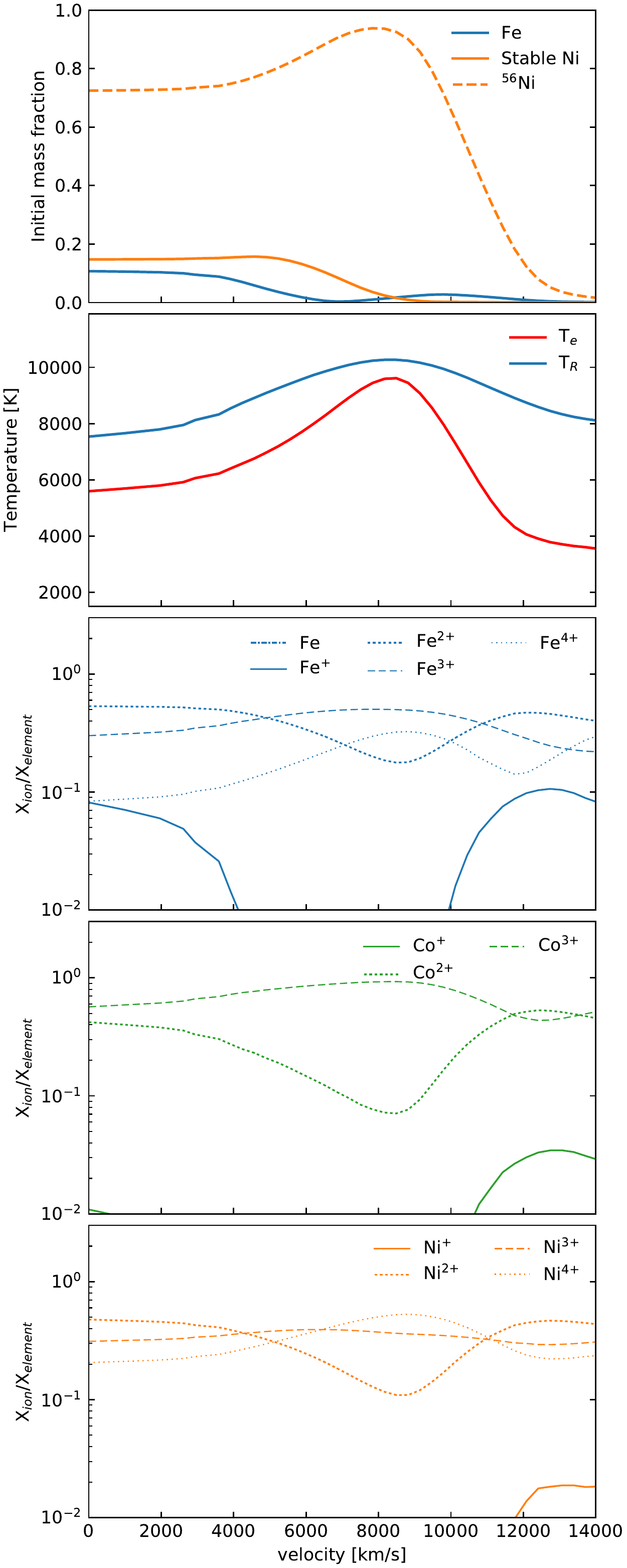}\end{center}
 \caption{Similar to \autoref{fig:330d-w7-estimators} but for the sub-\Mch detonation models at 330d: S0 (left) and with S5.5 (right). \label{fig:330d-subch-estimators}}
\end{figure*}

We have calculated \artis nebular spectra for both S0 and S5.5 for a range of epochs between 220 days and 360 days for both models.
\autoref{fig:330d-subch-estimators} shows
the temperatures and ionisation balance for both models at 330 days (i.e. the same epoch for comparison to W7 as shown in \autoref{fig:330d-w7-estimators}).
Compared to W7, the S0 and S5.5 models exhibit significantly higher ionisation, particularly in the core region, but also generally throughout the ejecta (compare bottom panels of Figures \ref{fig:330d-subch-estimators} and \ref{fig:330d-w7-estimators}).
Compared to the S0 model, the core of stable IGEs in the S5.5 model leads to lower electron temperatures and a lower ionisation state among Fe-group species at 330 days.
However, the S5.5 model is still hotter and more highly ionized than W7 throughout the core of the ejecta.

Figures \ref{fig:emissionspectra_s0} and \ref{fig:emissionspectra_s5p5} show the spectra of the \artis models for S0 and S5.5 at 330 days coloured by the emitting ion.
Despite the differences in composition and temperatures,
both models show broadly similar spectra to each other: dominated by
emission of Fe-peak elements.
However, there are clear differences that allow the models to be distinguished from W7.
In the optical, the most noticeable difference between W7 and the detonation models is in the emission of \ion{Ni}{ii}.
Both of the sub-\Mch models predict negligible emission by [\ion{Ni}{ii}] $\lambda\lambda$7378, 7412 and
the difference between the settling (S5.5) and the homogeneous (S0) model is not visible in this feature.
Due to the high ionisation state, the additional Ni abundance from increased stable \iso{58}{Ni} in S5.5 results in stronger [\ion{Ni}{III}] features than the S0 model.
In addition, the [\ion{Fe}{II}] features are much stronger in our W7 calculation than in the detonation models (a consequence of the lower ionisation state) and we note that absorption by \ion{Fe}{I} is much less important in the detonation models than in W7 (see Section \ref{sec:w7comparison}).

In the near-IR, only the W7 \artis calculations produce a clear [\ion{Ni}{ii}] 1.939 $\mu$m emission feature.
This feature has been discussed in the context of synthetic spectra for models with stable Ni by \citet{Blondin:2018dm} and \citet{Wilk:2018ki}, and is a particularly useful signature of Ni since it is relatively unblended \citep{Dhawan:2018wx,Flors:2019ek} (see right panels of Figures \ref{fig:emissionspectra_s0} and \ref{fig:emissionspectra_s5p5}).
Near-IR emission features from [\ion{Fe}{ii}] are virtually absent in the S0 model, while in S5.5 they become significant and exhibit a similar distribution to W7 (although they are much weaker than for W7).

% Aside from the overall shift in ionisation and presence of [\ion{Ni}{ii}] features, one prominent difference between W7 and our sub-\Mch calculations is the feature at $\sim$ 11\,800 \angstrom.
% In our simulations, this feature is formed by \ion{Fe}{i} fluorescence (z$^5$D$^{\mathrm{o}}\rightarrow\,$a$^5$P$^{\mathrm{e}}$ and z$^3$D$^{\mathrm{o}}\rightarrow\,$b$^3$P$^{\mathrm{e}}$) of mostly \ion{Fe}{ii} emission.
% This feature is absent in the sub-\Mch detonation spectra, consistent with the low neutral Fe abundance.

\subsection{Comparison to observations}

\begin{figure*}
 \begin{center}\includegraphics[width=0.5\textwidth]{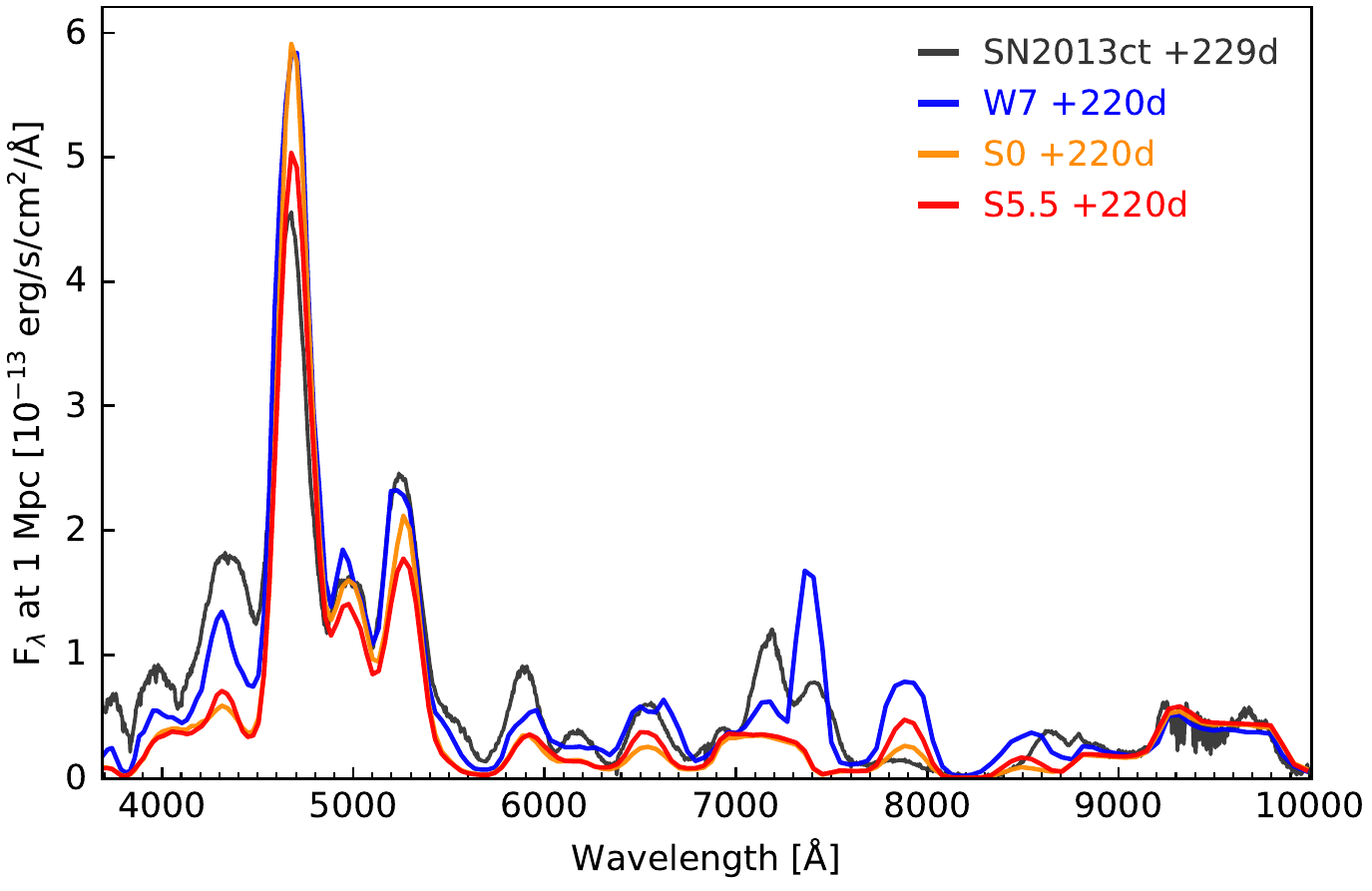}\includegraphics[width=0.5\textwidth]{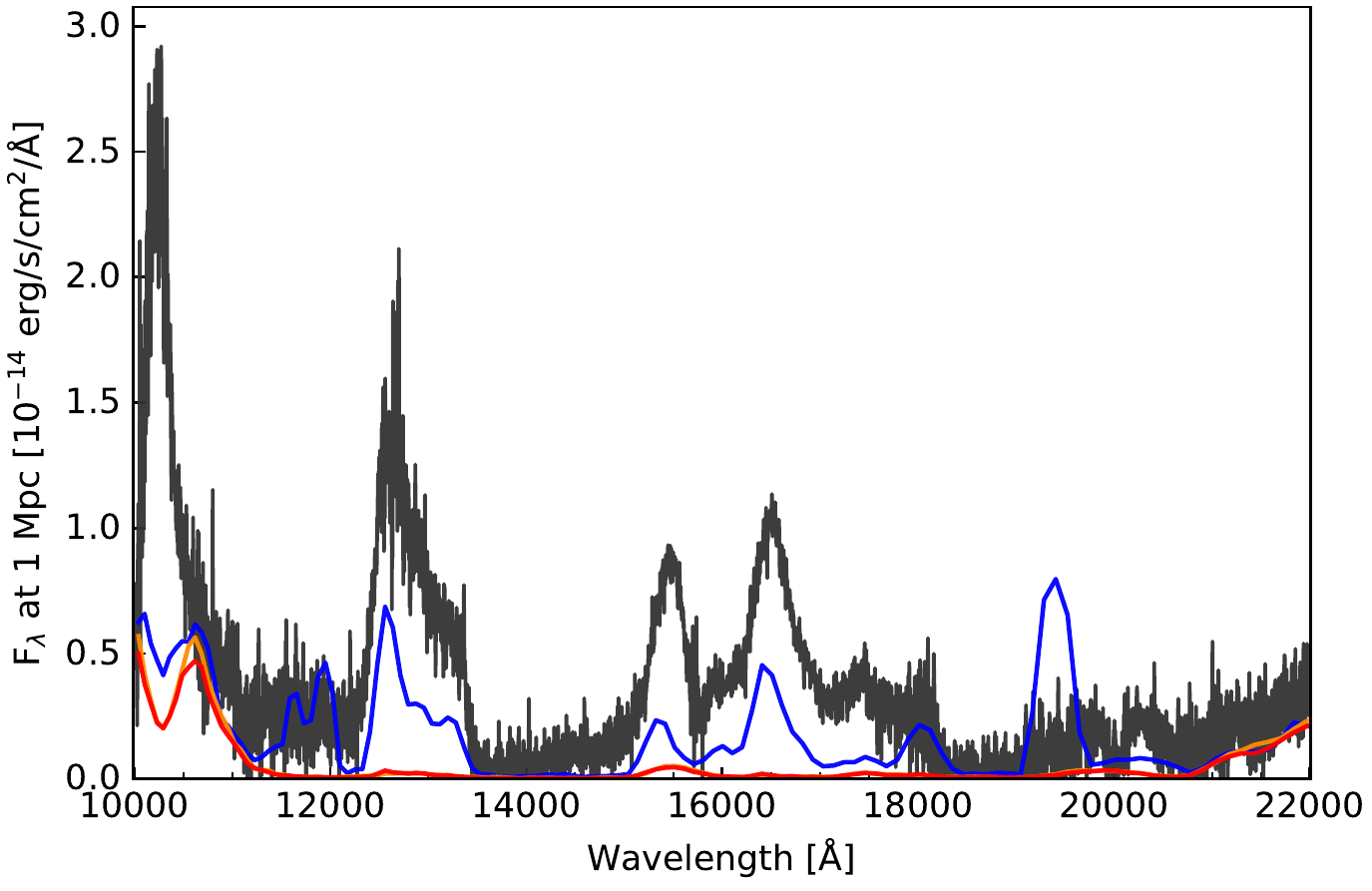}\end{center}
\caption{Nebular spectra of W7 and sub-\Mch \artis models compared to observed spectra of SN2013ct at 229d \citep{Maguire:2016jt}. \artis models at 220d in the optical (left) and the near-infrared (right). Note the differing scales on left and right panels.
\label{fig:specwithobs_220d}} \end{figure*}

\begin{figure*}
 \begin{center}\includegraphics[width=0.5\textwidth]{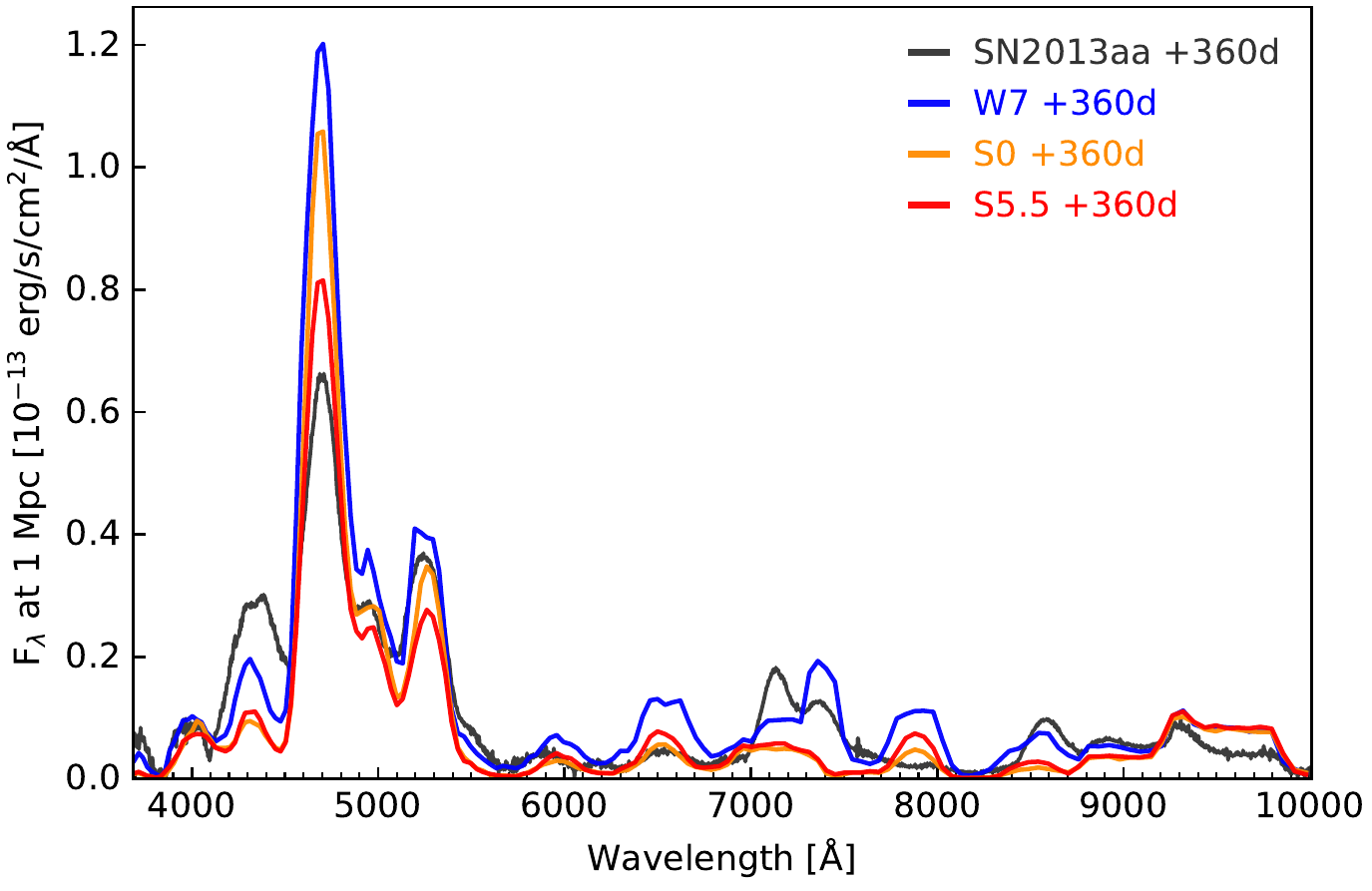}\includegraphics[width=0.5\textwidth]{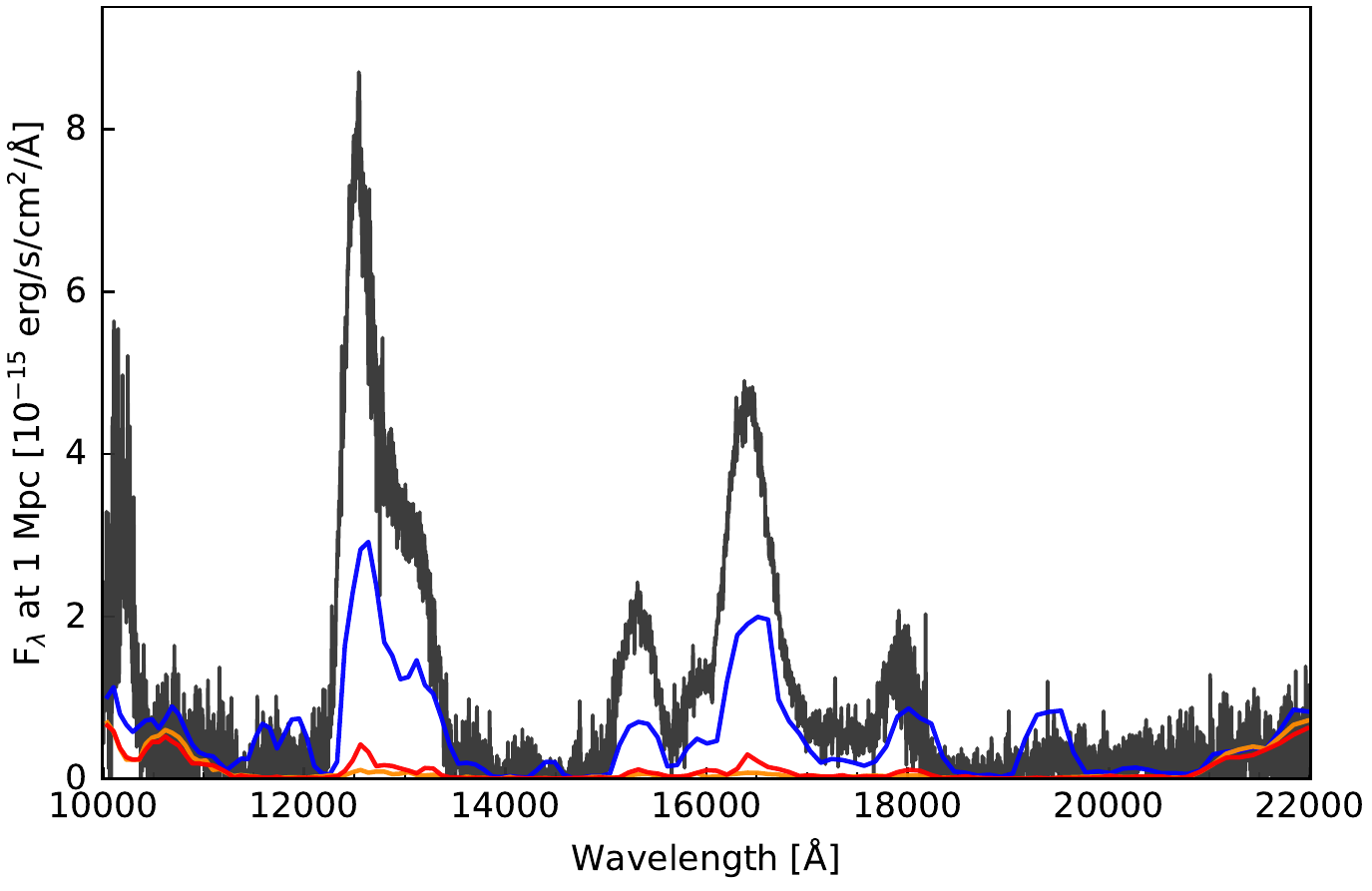}\end{center}
 \caption{Nebular spectra of W7 and sub-\Mch \artis models compared to observations of SN2013aa at 360d \citep{Maguire:2016jt}. \artis models at 360d in the optical (left) and the near-infrared. Note the differing scales on left and right panels.
\label{fig:specwithobs_360d}} \end{figure*}

Figures \ref{fig:specwithobs_220d} and \ref{fig:specwithobs_360d} show our synthetic spectra for the W7 and two sub-\Mch models compared to observed spectra at two epochs (220 and 360 days after explosion) for wavelengths from the optical to the near-IR.
The objects selected for comparison (SN2013ct and SN2013aa) have been chosen from the observed sample of \citet{Maguire:2016jt} for their simultaneous coverage in optical and near-IR at epochs roughly matching those of our \artis simulations.
Both objects are spectroscopically normal SNe~Ia \citep{Maguire:2013bd,Maguire:2016jt}, and the observed spectra have been corrected for redshift, reddening, and distance using the same values as \citet{Maguire:2016jt}.
Since the observed SN2013ct spectrum has an uncertain flux calibration, this spectrum has also been scaled by a constant normalisation factor for ease of comparison between features and the models.
The SN2013aa spectra are shown on the (calibrated) absolute flux scale.

In the optical region of the spectrum at 220 days (top left of \autoref{fig:specwithobs_220d}), the theoretical models produce a similar complex of [\ion{Fe}{ii}] and [\ion{Fe}{iii}] features around 4000 -- 5300~\AA.
In general the match of these features with the observation shown is reasonable, but there are clear discrepancies.
For example around 4300~\AA~where all of the models fail to reproduce the strength of the clear emission peak as seen in the data.
We note that this region was also where some discrepancies between \sumo and \artis manifest (see Section \ref{sec:w7comparison}), possibly suggesting systematic shortcomings in the modelling.
In agreement with \cite{Liu:1997hs} our W7 model calculation produces an optical [\ion{Ni}{ii}] $\lambda\lambda$7378, 7412 emission that is significantly too strong to match the observed spectrum.
The sub-\Mch detonation models on the other hand, predict negligible emission from [\ion{Ni}{ii}].
We note that all our models fail to reproduce the [\ion{Fe}{ii}] peak around 7200~\AA~at this epoch.
At the later epoch considered (+360 days), the agreement between the models and the observation of SN2013aa is of comparable quality to that between the model and SN2013ct around +220 days.
Again, the sub-\Mch models produce no significant [\ion{Ni}{ii}] feature and there are significant discrepancies around [\ion{Fe}{ii}] 4300~\AA, and the main [\ion{Fe}{iii}] peak is somewhat overproduced.

The right panels of \autoref{fig:specwithobs_220d} compare the near-IR spectra of the models to observations (same comparison supernovae and with consistent flux scaling relative to the optical).
Here we note that the [\ion{Fe}{ii}] emission from the sub-\Mch models is substantially too weak compared to the observed spectrum for the 220d SN2013ct spectrum.
In contrast, the W7 [\ion{Fe}{ii}] emission is also too weak but matches this spectrum better, except that the [\ion{Ni}{ii}] feature at 1.939 $\mu$m is far too strong in this calculation.

% The $\sim$ 11000 \angstrom feature (mostly produced by \ion{Fe}{i} fluorescence of [\ion{Fe}{ii}] emission) in our W7 spectra is loosely consistent with the observed spectra (SN2013ct and SN2013aa) although its presence in the observations is not conclusive.

The weakness of [\ion{Fe}{ii}] near-IR emission in the sub-\Mch detonations can most likely be attributed to the degree of ionisation being too high in the calculations. Despite the key role played by non-thermal ionisation in the overall ionisation structure, we find that photoionisation of the singly-ionised species is very important throughout much of ejecta -- indeed for Fe$^{{+}}$ we find that the photoionisation rate is comparable to or exceeds the non-thermal ionisation rate throughout the ejecta in the sub-\Mch models.

The origin of these high photoionisation rates is ultimately the photons (bound-free and free-free) produced by the high ionisation species, particularly Fe$^{3+}$ and Fe$^{4+}$. As first considered by \citet{Axelrod:1980vk} for the case of Type Ia supernovae in their nebular phase, recombination photons emitted by high Fe-group ions can easily photoionise lower ions (and neutral species).
The precise amount of `recycling' of recombination photons is extremely sensitive to the atomic data, ejecta structure, and non-thermal ionisation rates but plays an important role in our simulations.

The over-ionisation of sub-\Mch models was already found by \citet{RuizLapuente:1996hx} and is supported by other more recent studies.
In particular, our findings are similar to \citet{Mazzali:2015ey} and \citet{Wilk:2018ki}: both those studies found that over-ionisation is a challenge for modelling nebular spectra for several different scenarios, particularly sub-\Mch models.
Possible solutions to over-ionisation proposed include invoking a relatively large mass of stable Fe in the context of Chandrasekhar mass models \citep[i.e. \iso{54}{Fe}, see][]{Mazzali:2015ey} and/or potential clumping/inhomogeneities in the ejecta \citep{Wilk:2018ki,Wilk:2019vp}.
We note, however, that at the later epoch we consider the level of discrepancy between the sub-\Mch models and the observation of SN2013aa is smaller, albeit acting in the same sense.
Also the ratios of the spectral features are predicted to change as the ejecta conditions evolve.
This leads to complex changes in the predicted [\ion{Fe}{ii}] emission but also affects the [\ion{Ni}{ii}] 1.939 $\mu$m feature, which becomes much weaker at 360 days, relative to the other near-IR emission features, in the W7 model.

In summary, our results confirm that photoionisation is an important process, even at nebular phases.
This highlights the need for ongoing efforts to improve and expand atomic data for the iron-peak elements.

\section{Summary}
We have extended the validity of our three-dimensional Monte Carlo radiative transfer code \artis for SNe~Ia up to hundreds of days after maximum light.
We have achieved this by modelling the relevant physics in the nebular phase, including a detailed treatment of non-thermal electrons (the dominant source of ionisation in this phase), a new atomic dataset with forbidden transitions, a non-LTE population and ionisation solver, and a detailed non-LTE radiation field model.
As a test case, we have compared our results for the well-known W7 model with those of \sumo and found generally good agreement, with some discrepancies attributed to differences in the atomic processes.

We investigated the influence of gravitational-settling in the nebular-phase spectra of a sub-\Mch detonation model, and found that \iso{22}{Ne} settling is relevant to quantitative analysis and interpretation of SNe~Ia spectra in the context of sub-\Mch models.
In particular, settling enhances the predicted [\ion{Ni}{iii}] features and slightly lowers the ionisation balance overall.
In agreement with previous studies \citep{Mazzali:2015ey,Wilk:2018ki}, we do find that over-ionisation is a major issue for sub-\Mch models, and represents a considerably larger obstacle for reconciling this scenario with observations than any under- or over-prediction of the \iso{58}{Ni} abundance.
However, our comparisons for the two epochs considered here suggest that the magnitude of this discrepancy varies, indicating the potential value for future studies in which time series of nebular phase spectra can be consistently modelled.

In future studies, we intend to apply our radiative transfer method to three-dimensional explosion models, such as the DDT models of \citet{Seitenzahl:2013fz}.
These theoretical nebular spectra will help to resolve the degeneracy between candidate SNe~Ia scenarios and allow us to more fully quantify the observable signatures of the 3D structure of nucleosynthesis ash as predicted by modern hydrodynamical simulations.

\section*{Acknowledgments}
The authors thank Anders Jerkstrand for useful comments and discussion, and D. John Hillier for making the \cmfgen atomic data set publicly available.

SS, LS, and KM acknowledge support from STFC through grant, ST/P000312/1.
KM is supported by STFC through an Ernest Rutherford Fellowship (ST/M005348/1).
MB acknowledges support from the Swedish Research Council (Vetenskapsr\aa det) and the Swedish National Space Board.

IRS was supported by the Australian Research Council Grant FT160100028
and acknowledges Research Technology Services at UNSW for supporting this project with computing and storage resources.

The work of AM, MK and FKR is supported by the Klaus Tschira Foundation, by the Sonderforschungsbereich SFB 881 ``The Milky Way System'' of the German Research Foundation (DFG) and through computational resources at Forschungszentrum J{\"u}lich via project HMU14.

This research was undertaken with the assistance of resources, in particular the UNSW Merit Allocation Scheme and the Flagship Allocation Scheme, from the National Computational Infrastructure (NCI), which is supported by the Australian Government.

This work was performed using the Cambridge Service for Data Driven Discovery (CSD3), part of which is operated by the University of Cambridge Research Computing on behalf of the STFC DiRAC HPC Facility (www.dirac.ac.uk). The DiRAC component of CSD3 was funded by BEIS capital funding via STFC capital grants ST/P002307/1 and ST/R002452/1 and STFC operations grant ST/R00689X/1. DiRAC is part of the National e-Infrastructure.

This research has made use of NASA's Astrophysics Data System.
Figures in this work have been generated with the \textsc{matplotlib} package \citep{Hunter:2007ih}.

\bibliographystyle{mnras}
\bibliography{references}

\bsp

\label{lastpage}
\end{document}